\pdfoutput=1
\documentclass[journal]{elsarticle}
\usepackage{lineno,hyperref}
\usepackage[paperwidth=18.4cm,paperheight=26cm,top=1.5cm,bottom=2cm,right=2cm]{geometry}
\usepackage{amssymb}
\usepackage{amsmath}
\usepackage{hyperref} 
\hypersetup{
colorlinks=true,
linkcolor=blue,
filecolor=blue,
urlcolor==blue,
citecolor=blue,
}
\biboptions{sort&compress} 
\usepackage{color}
\usepackage{lineno}
\usepackage{setspace}
\usepackage[noend]{algpseudocode}
\usepackage{graphicx}
\usepackage{float} 
\usepackage{subfig}
\usepackage{algorithmicx,algorithm}
\usepackage{multirow}
\usepackage{bbm}
\def\degree{${}^{\circ}$}

\journal{arXiv}

\UseRawInputEncoding
\begin{document}

\begin{frontmatter}

\title{Fast simulation of airfoil flow field via deep neural network}

\cortext[mycorrespondingauthor]{Corresponding author}

\author[nwpu_address]{Kuijun Zuo}

\author[nwpu_address]{Zhengyin Ye}

\author[nwpu_address]{Shuhui Bu}

\author[cardc_address]{Xianxu Yuan \corref{mycorrespondingauthor}}
\ead{yuanxianxu@cardc.cn}

\author[nwpu_address]{Weiwei Zhang \corref{mycorrespondingauthor}}
\ead{aeroelastic@nwpu.edu.cn}

\address[nwpu_address]{School of Aeronautics, Northwestern Polytechnical University, Xi'an, 710072, China}

\address[cardc_address]{State Key Laboratory of Aerodynamics, China Aerodynamics Research and Development Center, Mian'yang, Si'chuan 621000, China}

\begin{abstract}
{ \indent 
Computational Fluid Dynamics (CFD) has become an indispensable tool in the optimization design, and evaluation of aircraft aerodynamics.
However, solving the Navier-Stokes (NS) equations is a time-consuming, memory demanding and computationally expensive task.
Artificial intelligence offers a promising avenue for flow field solving.
In this work, we propose a novel deep learning framework for rapidly reconstructing airfoil flow fields.
Channel attention and spatial attention  modules are utilized in the downsampling stage of the UNet to enhance the feature learning capabilities of the deep learning model.
Additionally, integrating the predicted flow field values generated by the deep learning model into the NS equation solver validates the credibility of the flow field prediction results.
The NACA series airfoils were used to validate the prediction accuracy and generalization of the deep learning model.
The experimental results represent the deep learning model achieving flow field prediction speeds three orders of magnitude faster than CFD solver.
Furthermore, the CFD solver integrated with deep learning model demonstrates a threefold acceleration compared to CFD solver.
By extensively mining historical flow field data, an efficient solution is derived for the rapid simulation of aircraft flow fields. 
%

}
\end{abstract}


\end{frontmatter}

\section{Introduction}
With the development of mathematical theories \cite{ershkov2021towards}, data science \cite{pathak2023big} and high-performance computing \cite{barrachina2023reformulating}, CFD plays a crucial role in aerospace \cite{basri2023computational}, energy and power science \cite{liu2023cfd, vimalakanthan2023computational}, and transportation industries \cite{bandi2023cfd}. 
It has gradually evolved into the foundation of large-scale equipment digital engineering and a key supporting tool \cite{jiangtao2023advances}. 
The CFD vision 2023 road map released by NASA highlights that CFD has transformed aircraft design methodologies, enhanced the capability to design complex aircraft, and reduced the design cycle for aircraft \cite{cary2021cfd}.
CFD technology primarily acquires flow field information by solving highly complex nonlinear NS equations.
However, the high-fidelity computational models, such as direct numerical simulation (DNS), requires a substantial amount of computational resources is becoming increasingly prohibitive \cite{gruber2022comparison}.

Since reduced order model (ROM) can approximate large-scale systems with lower computational costs, they can serve as an alternative to overcoming the trade-off between high-fidelity simulations and high computational cost \cite{dar2023artificial}.
In the field of scientific computing, the most commonly used reduced order models (ROMs) are the Proper Orthogonal Decomposition (POD) \cite{dowell1996eigenmode} and Dynamic Mode Decomposition (DMD) \cite{schmid2010dynamic} methods \cite{cao2019constrained, wu2018snapshot, schmid2011application, zhang2022unsteady}.
The successful implementation of dimensionality reduction in fluid dynamics depends on significant improvements in computer computational speed and memory capacity.
For the POD and DMD methods, capturing transient, intermittent, invariance, and multi-scale phenomena is challenging due to the inherent translation, rotation, and scaling characteristics of fluid \cite{kutz_2017}.

Inspired by the widespread success of machine learning in areas such as computer vision \cite{wang2023yolov7, zhang2023motrv2}, natural language processing \cite{hariri2023unlocking, tinn2023fine}, and autonomous driving \cite{teng2023motion, chitta2023transfuser}.
Considering extracting high-dimensional multi-scale features from a vast amount of fluid data using deep learning methods to find an alternative or improve existing expensive experimental and time-consuming iterative simulation tasks \cite{DURU2022105312}.
Some of the preliminary work already conducted has demonstrated the effectiveness of artificial intelligence approaches in accelerating scientific computing \cite{GRUBER2022114764, HU2022115128, MA2022115496, KISSAS2020112623, LI2023116299}.
In the field of aerodynamics, Leer et al. \cite{Leer2020FastFF} used a deep learning approach with multilayer perceptron (MLP) and radial-logarithmic filter mask (RLF) to predict incompressible laminar steady flow fields for various geometric shapes.
Sun et al. \cite{100060604} utilized deep learning methods to predict the characteristics of compressible flow for supersonic airfoils, such as lift, drag, and pitch coefficients, achieving a high level of prediction accuracy and efficiency.

Some pioneering work has shown that machine learning methods can be used for turbulence modeling \cite{lav2023coupled, WEATHERITT201622, grabe2023data, tenachi2023deep} and rapid flow field prediction under different flow conditions \cite{J061234, sekar2019fast, leer2021fast, LIU2023126425, LINKA2022115346}.
However, when dealing with large-scale flow field data, neural network models like MLP face challenges such as a large number of parameters and high training time costs.
Due to the weight sharing property of convolutional neural networks (CNN) \cite{zuo2022fast}, they can significantly reduce the number of model training parameters compared to MLP, and as a result, they are widely used in flow field prediction tasks \cite{duru2022deep, duru2021cnnfoil, guo2016convolutional, thuerey2020deep}.
For instance, Ribeiro et al. \cite{ribeiro2020deepcfd} proposed a deep neural network model based on convolutional neural networks, known as deepCFD, for predicting non-uniform steady laminar flow problem.
Chen and Nils \cite{chen2023towards} conducted research using the UNet deep neural network to address the inference problem of exact solutions for two-dimensional compressible Reynolds-averaged Navier-Stokes (RANS) flow over airfoils.
The UNet \cite{ronneberger2015u} neural network primarily consists of a topology composed of convolutional upsampling blocks and convolutional downsampling blocks, with feature fusion between the upsampling and downsampling layers facilitated through ``skip connection'' operations.
Both UNet++ \cite{zhou2019unet++} and UNet3+ \cite{huang2020unet} enhance the feature extraction capability of deep learning models by adding more densely skip connections. 
However, solely increasing the network's depth and dense connections may not be an optimal solution, as it adds to the memory cost of model training.
Human perception of large-scale information is primarily achieved by focusing on significant aspects while disregarding irrelevant details.
Hence, our proposed deep learning framework for rapid simulation of aircraft airfoils, FU-CBAM-Net, enhances the extraction of high-dimensional flow field nonlinear features by adding channel attention and spatial attention behind the UNet convolutional downsampling blocks.
FU-CBAM-Net applies channel attention and spatial attention operations to the input features separately, reducing attention to irrelevant noise. Furthermore, it performs element-wise multiplication between the attention maps and the output features from the previous convolutional layer, achieving adaptive feature refinement \cite{woo2018cbam}.
Despite the rapid development of machine learning-based physics simulations in recent years, researchers in the field still harbor skepticism regarding the accuracy and generalizability of machine learning methods \cite{durbin2018some}. Many researchers consider it as a black-box models that are difficult to understand and explain \cite{thuerey2020deep}.
Here, to enhance the credibility of the deep neural network's flow field predictions, the FU-CBAM-Net prediction values are coupled with the PHengLEI \cite{zhong2019phenglei} solver developed by the China Aerodynamics Research and Development Center (CARDC) to iteratively solve the NS equations.

The main contributions of this work are as follows:
\begin{itemize}
	\item We propose a novel deep convolutional attention network model for predicting flow field solutions of different airfoil shapes under varying operational conditions.
	\item By embedding the FU-CBAM-Net model into the PHengLEI solver, it not only enhances the credibility of FU-CBAM-Net flow field prediction results, but also accelerates the convergence speed of aerodynamic parameters during PHengLEI's solving process, reducing the number of solver iterations.
	\item We have made our code publicly available on \href{https://github.com/zuokuijun/FU-CBAM-Net}{GitHub} repository to enable other researchers to replicate our work.
\end{itemize}

The rest of this paper is organized as follows. 
Section \uppercase\expandafter{\romannumeral2} 
mainly describes the deep learning methods used for predicting airfoil flow fields.
Section \uppercase\expandafter{\romannumeral3} primarily discusses the airfoil flow field dataset.
Section \uppercase\expandafter{\romannumeral4} shows and discusses the results of the FU-CBAM-Net neural network model training and prediction. 
And the conclusion is given in Section \uppercase\expandafter{\romannumeral5}.

\section{Methodology}

\subsection{Overview}

\begin{figure*}[!h]
	\begin{center}
		\includegraphics[width=1 \linewidth]{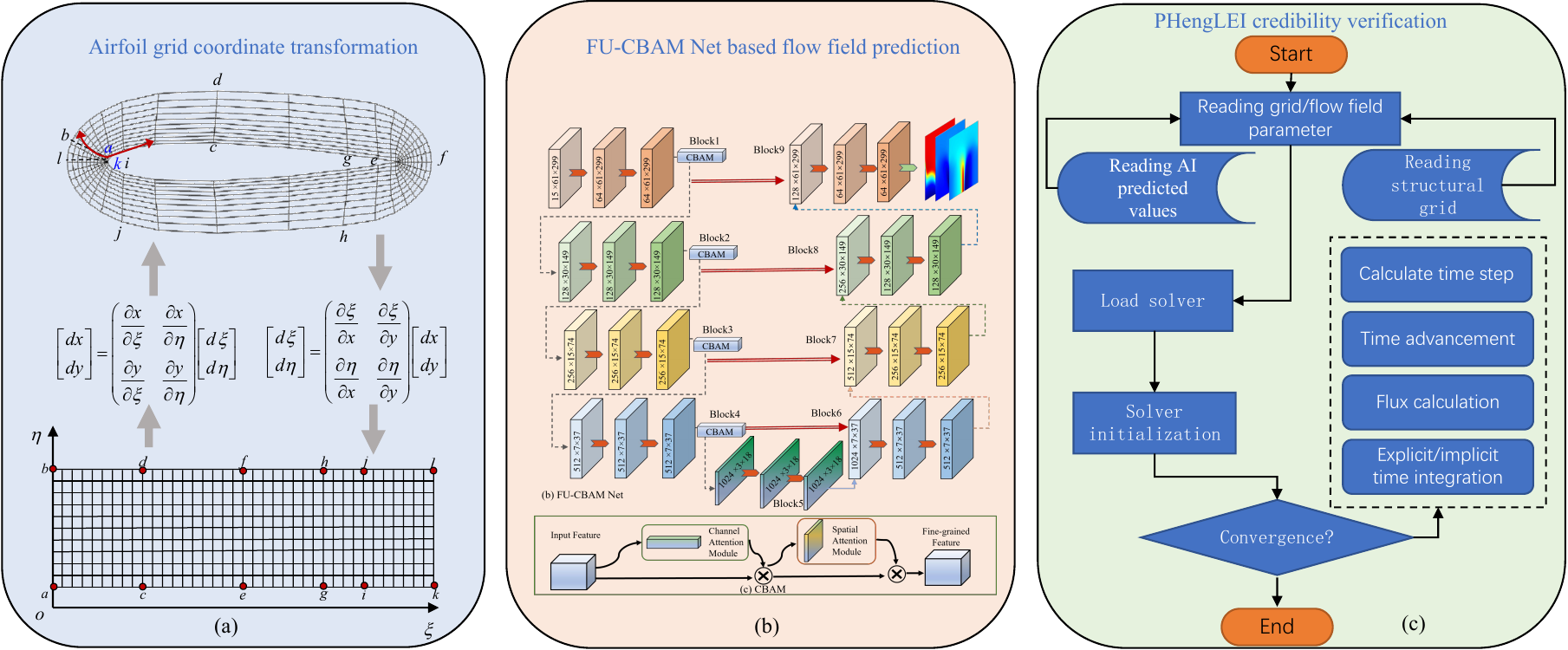}
	\end{center}  \vspace{-2mm}  
	\caption{{{Flowchart of flow field prediction and credibility assessment method coupling FU-CBAM Net with PHengLEI solver}
	}} \label{overview} 
\end{figure*}

The purpose of this study is to utilize machine learning methods to rapidly predict steady laminar flow fields for different airfoil shapes under various flow conditions, and to explore credible evaluation methods for flow field predictions obtained through deep learning techniques.
The algorithm's computational workflow is illustrated in Fig.\ref{overview}. 

The first part is airfoil flow field data preprocessing. 
As shown in Fig. \ref{overview}(a), for the purpose of facilitating subsequent feature extraction by FU-CBAM-Net, we consider transforming non-uniform grid data in the physical coordinate into uniform grid data in the computational coordinate. 
Using a series of NACA airfoil data, the PHengLEI solver was employed to obtain laminar flow field solutions for airfoils at Mach number 0.2, with angles of attack ranging from 0\degree\ to 5\degree, and Reynolds numbers from 1000 to 2000. For a detailed description of the flow field data, please refer to Section 3.

In accordance with the illustration in Fig. \ref{overview} (b), the improved UNet neural network, FU-CBAM-Net, is used for the task of rapid flow field prediction. 
The input to the deep neural network consists of flow field coordinates and physical parameters that represent the flow field state, such as Reynolds number (Re) and angle of attack (AOA).
The output includes velocity fields, pressure fields, and corresponding gradient information for the respective airfoil. 
For a detailed explanation of the working principles of FU-CBAM-Net and the specifics of its input and output parameters, please refer to Section \ref{Geometric information encoding} and Section \ref{fu-cbam-net}.

Most existing deep learning models for end-to-end flow field prediction tasks are predominantly black-box models.
In fields related to aerodynamic optimization design and the like, although the introduction of artificial intelligence has accelerated the speed of flow field resolution and improved prediction accuracy significantly, the inherent lack of interpretability in deep learning methods has left researchers in these areas somewhat skeptical about the practical engineering applications of such approaches.
As shown in Fig. \ref{overview}(c), to further enhance the credibility of deep learning flow field predictions, the predicted flow field from the FU-CBAM model is fed into the PHengLEI solver to solve the Navier-Stokes equations until convergence is achieved.
The flow field obtained through CFD techniques is widely accepted by researchers in the relevant field. 
For related test results, please refer to the fourth section.

\subsection{Geometric information encoding} \label{Geometric information encoding}

To accurately capture the flow field information in the boundary layer region of the airfoil, the geometry and flow information of the airfoil are transformed from Cartesian coordinates $(x,y)$ to curvilinear coordinates $(\xi, \eta)$ through univalent transformation:

\begin{equation}
\left[\begin{array}{c}
x \\
y
\end{array}\right]_{i, j}=\left[\begin{array}{c}
x_0 \\
y_0
\end{array}\right]_{i, 1}+\int\left[\begin{array}{c}
d x \\
d y
\end{array}\right] \label{eq:1}
\end{equation}

In Eq. \eqref{eq:1}, $x_0$ and $y_0$ represent the coordinate information of the given airfoil surface. 
$i$ and $j$ are indices that indicate different directions within the mesh.
Furthermore, we have:

\begin{equation} \label{eq:2}
	\left[\begin{array}{c}
	d x \\
	d y
	\end{array}\right]=\left[\begin{array}{cc}
	\frac{\partial x}{\partial \xi} & \frac{\partial x}{\partial \eta} \\
	\frac{\partial y}{\partial \xi} & \frac{\partial y}{\partial \eta}
	\end{array}\right]\left[\begin{array}{c}
	d \xi \\
	d \eta
	\end{array}\right] 
\end{equation}
and
\begin{equation} \label{eq:3}
	\left[\begin{array}{c}
	d \xi \\
	d \eta
	\end{array}\right]=\left[\begin{array}{cc}
	\frac{\partial \xi}{\partial x} & \frac{\partial \xi}{\partial y} \\
	\frac{\partial \eta}{\partial x} & \frac{\partial \eta}{\partial y}
	\end{array}\right]\left[\begin{array}{l}
	d x \\
	d y
	\end{array}\right] 
\end{equation}

As derived from Eq. \eqref{eq:2}:

\begin{equation}
	\left[\begin{array}{c}
	\mathrm{d} \xi \\
	\mathrm{d} \eta
	\end{array}\right]=\left[\begin{array}{ll}
	\frac{\partial x}{\partial \xi} & \frac{\partial x}{\partial \eta} \\
	\frac{\partial y}{\partial \xi} & \frac{\partial y}{\partial \eta}
	\end{array}\right]^{-1}\left[\begin{array}{l}
	\mathrm{d} x \\
	\mathrm{d} y
	\end{array}\right] \label{eq:4}
\end{equation}

Combining Eq. \eqref{eq:3} and Eq. \eqref{eq:4}, we obtain:

\begin{equation}
	\left[\begin{array}{cc}
	\frac{\partial \xi}{\partial x} & \frac{\partial \xi}{\partial y} \\
	\frac{\partial \eta}{\partial x} & \frac{\partial \eta}{\partial y}
	\end{array}\right]
	=
    \frac{1}{J}\left[\begin{array}{cc}
    \frac{\partial y}{\partial \eta} & -\frac{\partial x}{\partial \eta} \\
    -\frac{\partial y}{\partial \xi} & \frac{\partial x}{\partial \xi}
    \end{array}\right]
\end{equation}

where

\begin{equation}
	 J = \left|\begin{array}{cc}
	 \frac{\partial x}{\partial \xi} & -\frac{\partial x}{\partial \eta} \\
	 -\frac{\partial y}{\partial \xi} & \frac{\partial y}{\partial \eta}
	 \end{array}\right|
\end{equation}

Here, $J$ denotes the Jacobian matrix. 
Hence, the metric can be expressed as:
\begin{equation}
	\frac{\partial \xi}{\partial x} = \frac{1}{J}\frac{\partial y}{\partial \eta}, \quad
	\frac{\partial \xi}{\partial y} = -\frac{1}{J}\frac{\partial x}{\partial \eta}, \quad
	\frac{\partial \eta}{\partial x} = -\frac{1}{J}\frac{\partial y}{\partial \xi},  \quad
	\frac{\partial \eta}{\partial y} = \frac{1}{J}\frac{\partial x}{\partial \xi} 
\end{equation}

\begin{figure*}[!h]
	\begin{center}
		\includegraphics[width=1 \linewidth]{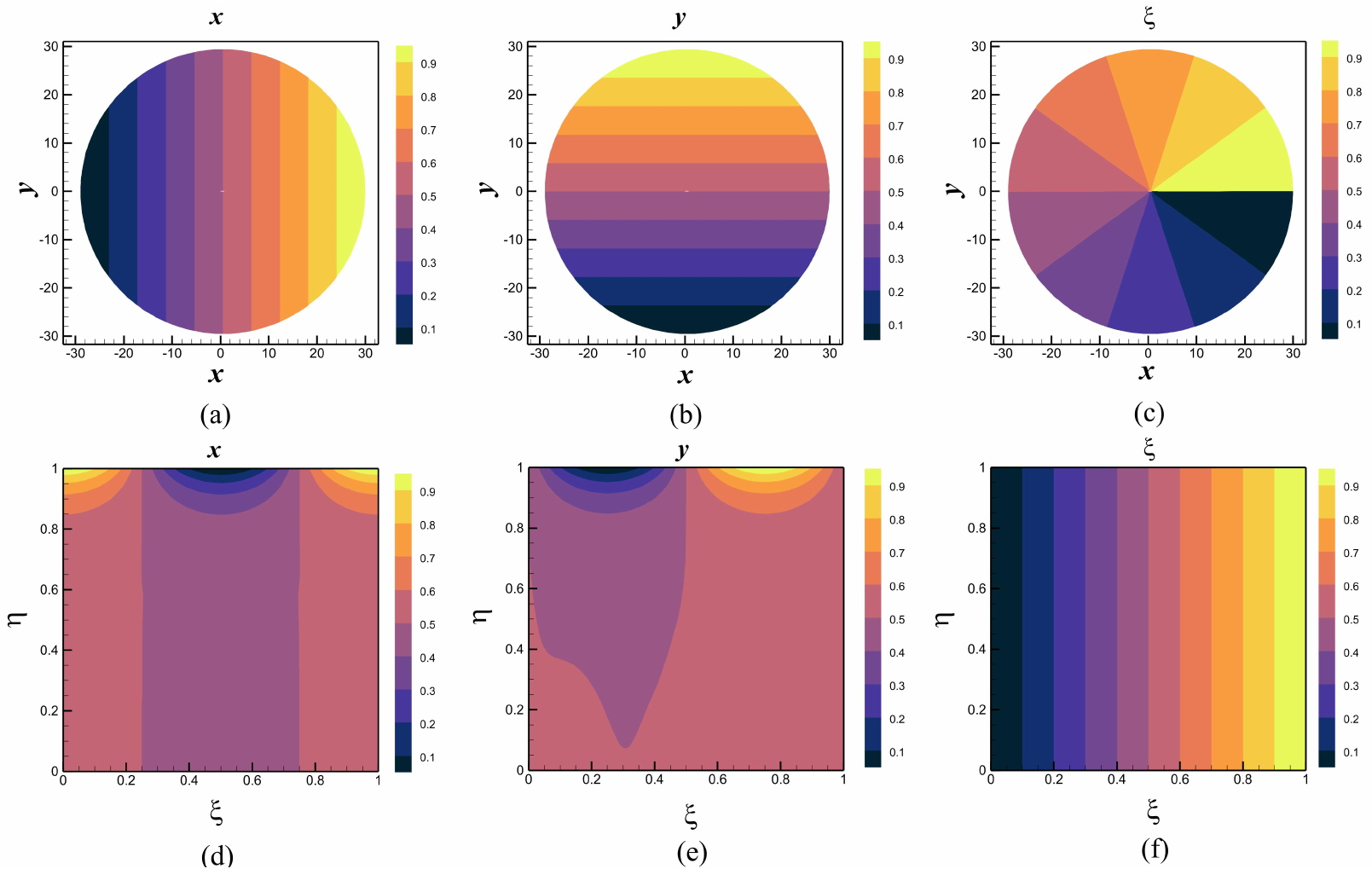}
	\end{center}  \vspace{-2mm}  
	\caption{{{Flow field coordinate map before ((a) Normalized $x$ coordinate, (b) Normalized $y$ coordinate, (c) $\xi$ map)and after ((d) Normalized $x$ coordinate, (e) Normalized $y$ coordinate, (f) $\xi$ map)mesh transformation.}
	}} \label{coordinate transformation 1} 
\end{figure*}

\begin{figure*}[!h]
	\begin{center}
		\includegraphics[width=1 \linewidth]{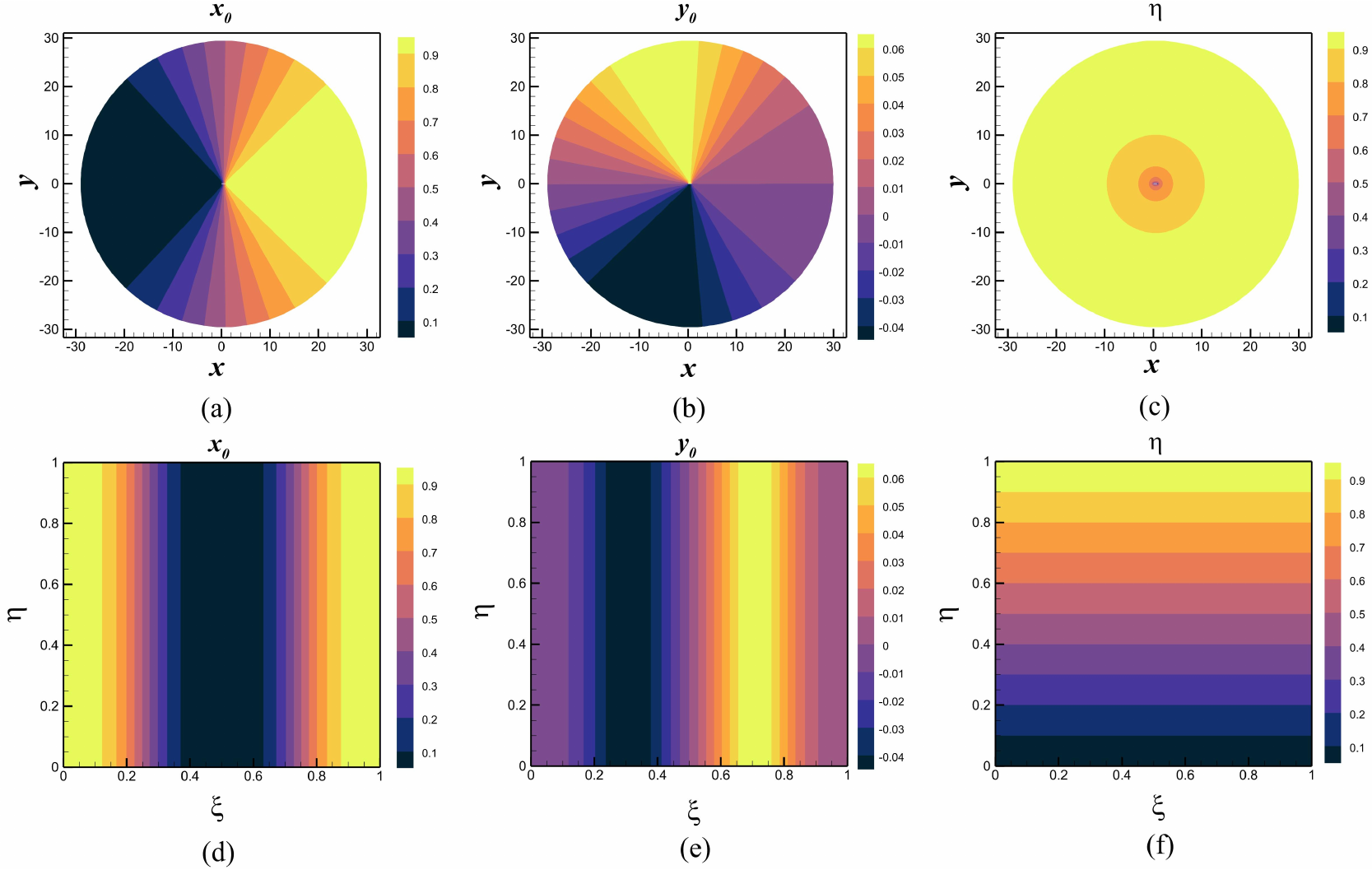}
	\end{center}  \vspace{-2mm}  
	\caption{{{Mapping diagrams of airfoil profile geometric coordinate before ((a) $x_0$, (b) $y_0$) and after ((d) $x_0$, (e) $y_0$)  transformation. Mapping diagrams of computed coordinate $\eta$ before ((c) $\eta$) and after ((f) $\eta$) mesh transformation.}
	}} \label{coordinate transformation 2} 
\end{figure*}

In the above formula, $\xi = (i-1)/(i_{max} - 1)$, $\eta = (j-1)/(j_{max} - 1)$.
$i_{max}$, $j_{max}$ correspond to the maximum number of mesh nodes in different directions.
Figure. \ref{coordinate transformation 1}, Fig. \ref{coordinate transformation 2}(c), Fig. \ref{coordinate transformation 2}(f) respectively present mapping diagrams of Cartesian coordinates and curvilinear coordinates before and after mesh transformation.
The flow field coordinates can reflect the spatial distribution of different physical quantities in various coordinate systems. Therefore, they are utilized as inputs for the neural network.
As shown in Fig. \ref{coordinate transformation 2}, the geometric coordinates of different airfoil shapes are also used as input parameters for the neural network.
The boundary layer region in the flow field often contains rich information and is a focal point in traditional CFD calculations. 
In contrast, far-field data does not require excessive attention.
According to reference \cite{deng2023prediction}, using the filter $M$, reweights the flow field coordinates to obtain two new parameters, $M_x$ and $M_y$, with the following calculation formula:

\begin{equation} \label{eq:8}
\left\{
\begin{aligned}
   \begin{split}
	M = e^{-|SDF|}, \Psi_M(x) = M \times x, \Psi_M(y) = M \times y, \\
	SDF(i, j)=\mathop{min} \limits_{(i^*, j^*) \in Z} |(i, j) - (i^*, j^*)| sign[f(i, j)]
	\end{split}
\end{aligned}
\right.
\end{equation}

\begin{figure*}[!h]
	\begin{center}
		\includegraphics[width=1 \linewidth]{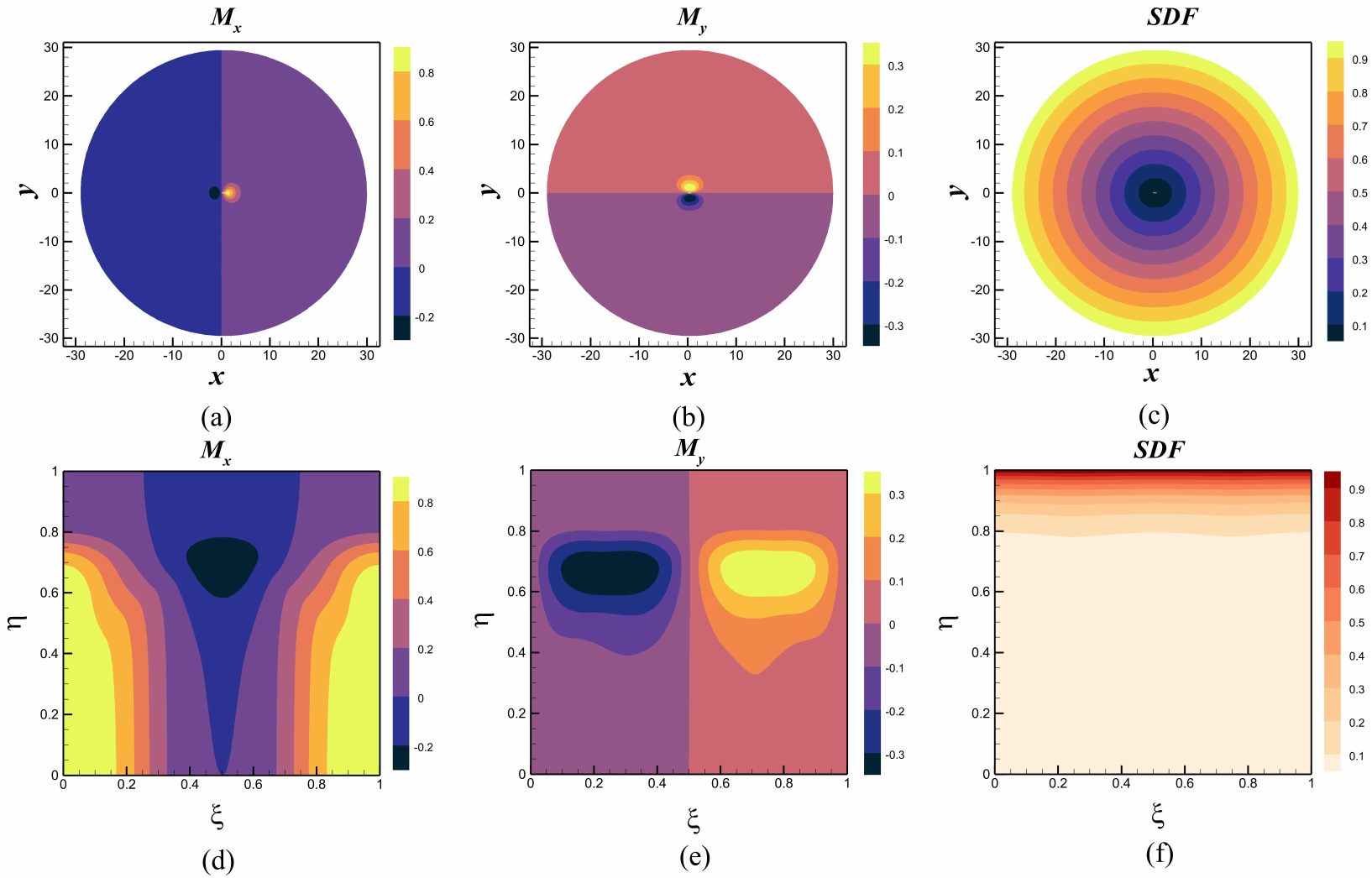}
	\end{center}  \vspace{-2mm}  
	\caption{{{Mapping diagrams for $M_x$, $M_y$, $SDF$ before ((a) $M_x$, (b) $M_y$, (c) $SDF$) and after ((d) $M_x$, (e) $M_y$, (f) $SDF$) mesh transformation.}
	}} \label{coordinate transformation 3} 
\end{figure*}

In Eq. \eqref{eq:8}, $SDF$ represents Signed Distance Field and $sign[f(i, j)]$ represents the symbolic function.
Here, the $SDF$ calculates the normal distance from each mesh point in the flow field to the airfoil profile.
Figure \ref{coordinate transformation 3} presents visual results for $M_x$, $M_y$, and $SDF$ in different coordinate systems.

\begin{figure*}[!h]
	\begin{center}
		\includegraphics[width=0.6 \linewidth]{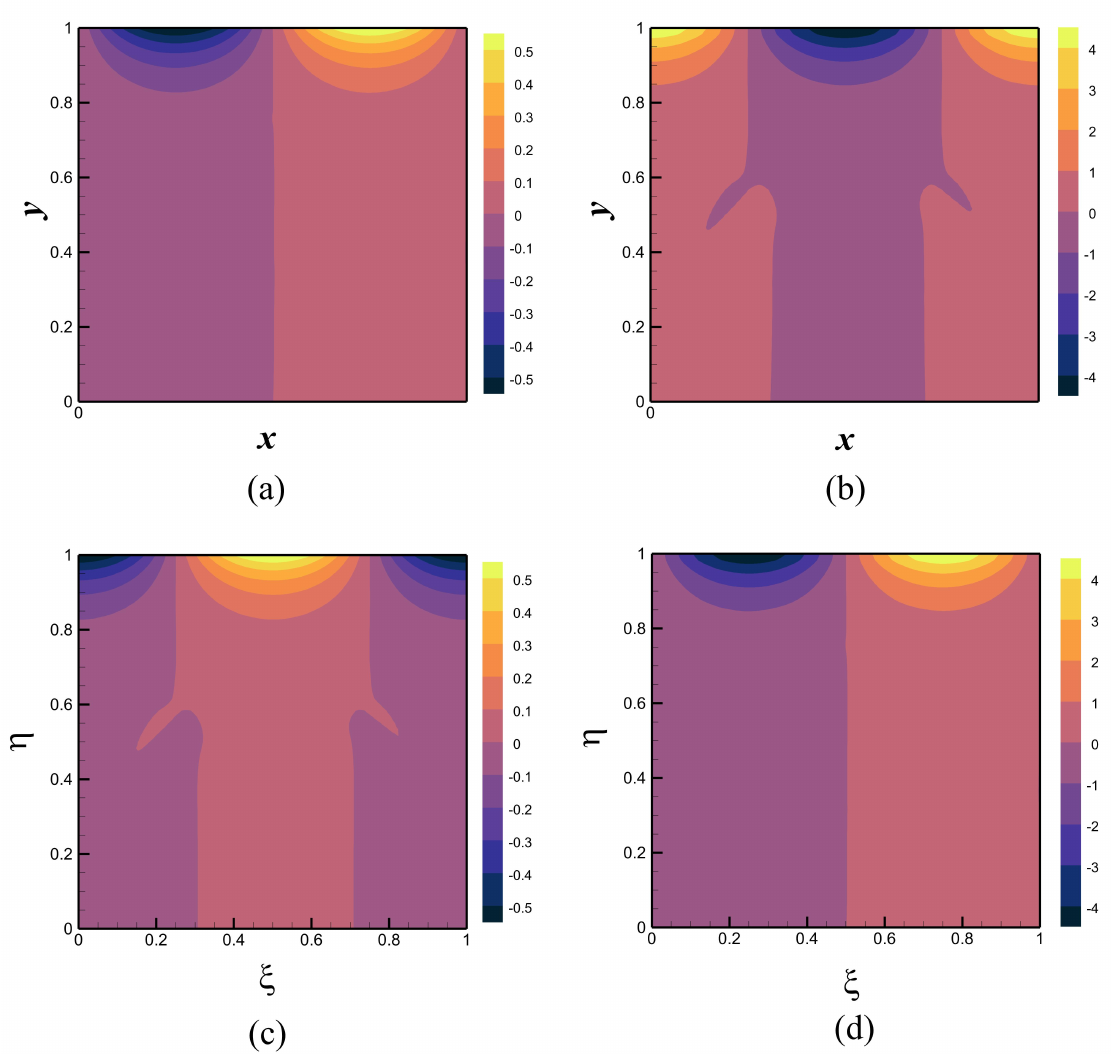}
	\end{center}  \vspace{-2mm}  
	\caption{{{Metrics in the governing equations. (a) $\frac{\partial x}{\partial \xi}$, (b) $\frac{\partial x}{\partial \eta}$, (c) $\frac{\partial y}{\partial \xi}$, (d) $\frac{\partial y}{\partial \eta}$.}
	}} \label{coordinate transformation 4} 
\end{figure*}

During the mesh transformation process, distortion may occur due to operations such as mesh stretching \cite{deng2023prediction}. 
To address this issue, the Jacobian matrix used in the mesh transformation is also employed as input information for the neural network. 
The Jacobian matrix parameters in computational coordinates are shown in Fig. \ref{coordinate transformation 4}.
Based on the analysis above, the neural network model FU-CBAM-Net has a total of fifteen input parameters, which are $x$, $y$, $x_0$, $y_0$, $\xi$, $\eta$, $SDF$, $M_x$, $M_y$, $AOA$, $Re$, $\frac{\partial x}{\partial \xi}$, $\frac{\partial x}{\partial \eta}$, $\frac{\partial y}{\partial \xi}$, $\frac{\partial y}{\partial \eta}$.

\subsection{UNet convolutional neural network} \label{Unet}

\begin{figure*}[!h]
	\begin{center}
		\includegraphics[width=1 \linewidth]{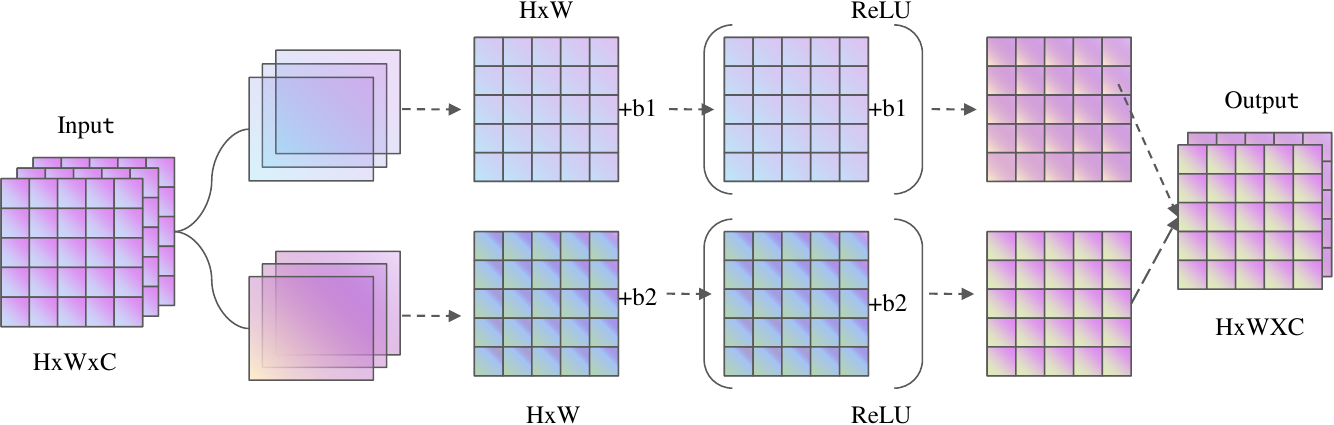}
	\end{center}  \vspace{-2mm}  
	\caption{{{Typical convolutional neural network architecture.}
	}} \label{cnn} 
\end{figure*}

The emergence of CNN has significantly improved the training speed of deep learning models compared to fully connected neural networks. 
As a fundamental module, CNN has been widely adopted in various fields, including computer vision \cite{yin2022cloud}, flow field prediction \cite{hu2023flow}, and aerodynamic optimization design \cite{sabater2022fast}.
Figure \ref{cnn} illustrates a standard convolutional neural network architecture. 
$X \in \mathbb{R}^{H \times W \times C}$ represents the input to the neural network, where $H$, $W$, and $C$ denote the height, width, and number of channels of the input feature maps, respectively.
CNN utilizes convolutional kernels to perform feature extraction by sliding windows over feature maps. 
The size of the output feature maps can be calculated using the following formula:

\begin{equation}
\left\{
\begin{aligned}
H_{out} &= \frac{H_{in} + 2 \times P_0 - D_0 \times (K_0 - 1) - 1}{S_0} + 1, \\
W_{out} &= \frac{W_{in} + 2 \times P_1 - D_1 \times (K_1 - 1) - 1}{S_1} + 1.
\end{aligned}
\right.   \label{CNN calculation formula}
\end{equation} 

In Eq. \eqref{CNN calculation formula},  $P_i$ represents the values to be padded on the four sides of the input features, typically set to 0.
$D_i$ denotes the spacing between kernel elements, which is set to 1 here.
$K_i$ represents the size of the kernel, and $S_i$ represents the stride of the kernel.
For more details, please refer to our previous work \cite{zuo2022fast}.

\begin{figure*}[!h]
	\begin{center}
		\includegraphics[width=1 \linewidth]{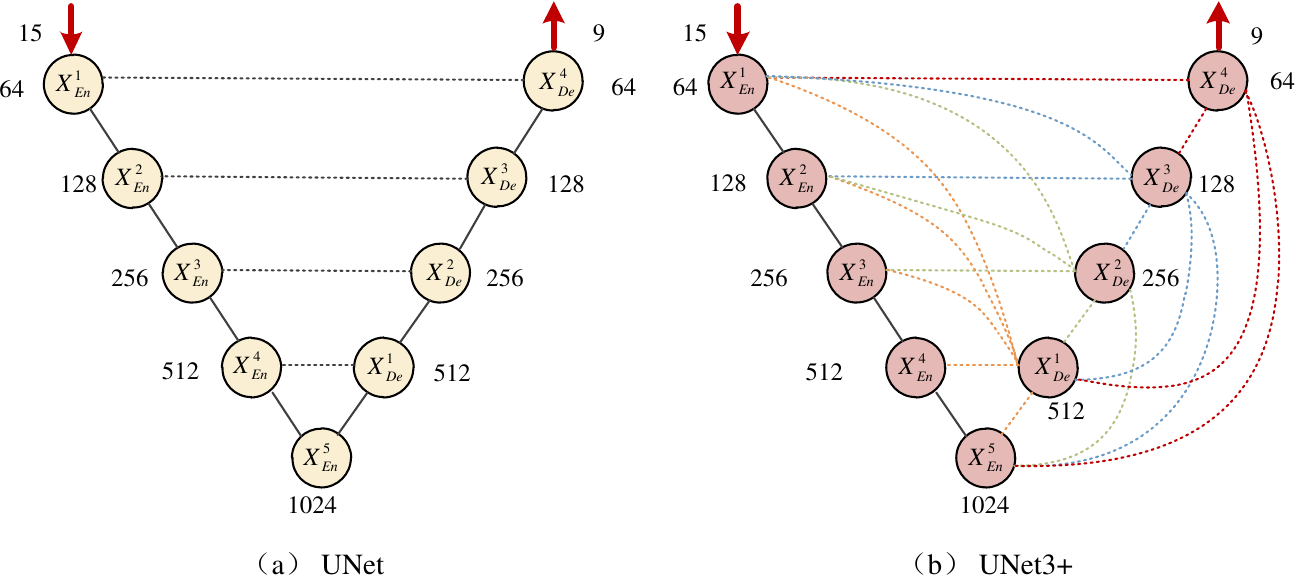}
	\end{center}  \vspace{-2mm}  
	\caption{{{UNet neural network and UNet3+ network architecture diagram.}
	}} \label{unet and unet3+} 
\end{figure*}
Deeper feature maps often have a larger receptive field. 
As shown in Fig. \ref{unet and unet3+}(a), Olaf et al. proposed a UNet neural network model based on a convolutional neural network architecture to address image segmentation problems in the medical field. 
Shallow convolutions are employed to extract low-level geometric features, while deep convolutions are used to extract high-level semantic features.
Additionally, to address the issue of information loss during the upsampling process of feature maps, a residual network structure is utilized to concatenate feature maps with the same number of channels, eliminating the interference of information loss on flow field prediction results.
However, the UNet neural network performs feature fusion only at the same feature scale. 
As depicted in Fig. \ref{unet and unet3+}(b), Huang and his team improved the model's prediction accuracy by using multiscale skip connections to integrate high-level semantic features from different scales along with low-level geometric features.
The feature maps after the $i-th$ decoder can be calculated using the following formula:
\begin{equation}
	X_{D e}^i= \begin{cases}X_{E n}^i, &i= N\\ \mathcal{H}\left(\left[\underbrace{\mathrm{C}\left(\mathcal{D}\left(X_{E n}^k\right)\right)_{k=1}^{i-1}, \mathrm{C}\left(X_{E n}^i\right)}_{\text {Scales: } 1^{t h} \sim i^{t h}}, \underbrace{\mathrm{C}\left(\mathcal{U}\left(X_{D e}^k\right)\right)_{k=1}^{i-1}}_{\text {Scales: }1^{t h} \sim i^{t h}} \right] \right) , &i= 1,...,N-1 \end{cases}
\end{equation}
where $\mathcal{D}(.)$ and $\mathcal{U}(.)$ represent the downsampling and upsampling operations, respectively.
The function $\mathrm{C}(.)$ represents the convolution operation.
$\mathcal{H}(.)$ denotes the operation of aggregating multiscale features.

\subsection{FU-CBAM-Net} \label{fu-cbam-net}

\begin{figure*}[!h]
	\begin{center}
		\includegraphics[width=1 \linewidth]{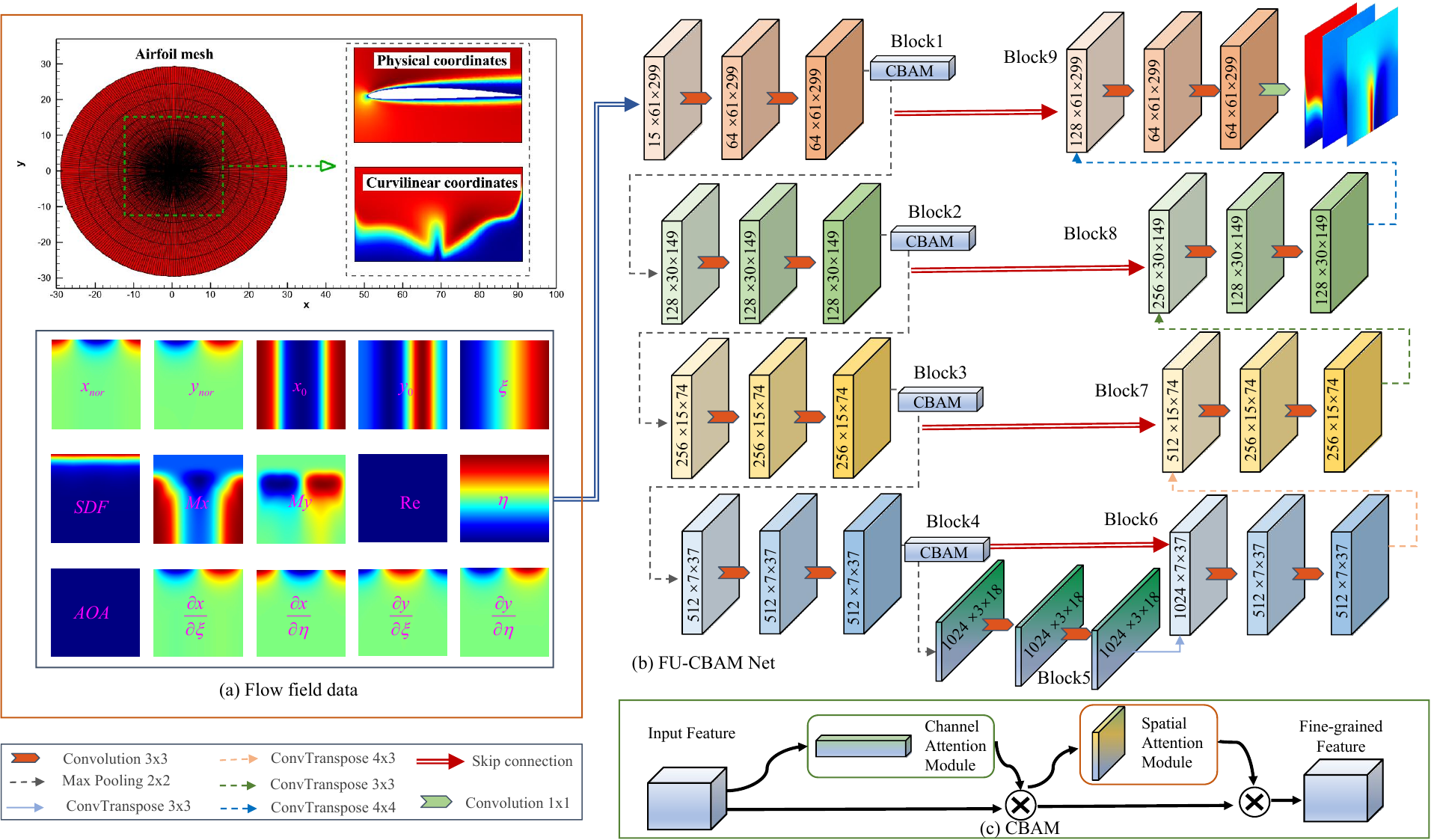}
	\end{center}  \vspace{-2mm}  
	\caption{{{An illustration of the network architecture for FU-CBAM-Net.}
	}} \label{FU-CBAM Net} 
\end{figure*}

When confronted with vast amounts of data, the human perceptual system tends to filter and concentrate on particular information zones.
By mimicking the human perceptual system's way of focusing on areas of interest, it is possible to enhance the efficiency and performance of computer vision and machine learning algorithms. 
This approach is known as the ``attention mechanism'' and has become an important component of modern deep learning.
In contrast to the processing approach of UNet and UNet3+ with skip connections, the attention mechanism can identify the most relevant regions and effectively enhance the feature fusion and representation capabilities of neural networks.
Therefore, we have adopted a feature fusion strategy different from UNet3+. 
An attention model in the form of a convolutional block is embedded during the downsampling process in UNet, using both spatial attention and channel attention strategies to enhance the model's predictive performance.
The FU-CBAM-Net network constructed as shown in Fig. \ref{FU-CBAM Net}.
Figure \ref{FU-CBAM Net} (a) illustrates the fifteen parameters fed into FU-CBAM-Net.
Each parameter matrix has a size of $61 \times 299$. 
These matrices are concatenated along the channels to form a tensor with 15 channels, which serves as the input for the neural network.
Figure \ref{FU-CBAM Net} (b) presents the changes in the dimensions of the tensor as it passes through each layer of the FU-CBAM-Net when a tensor of size $15 \times 61 \times 299$ is input. 
More detailed about FU-CBAM-Net architecture, please refer to Tbl. \ref{FU-CBAM_enecoder} and Tbl. \ref{FU-CBAM_decoder}.
Figure \ref{FU-CBAM Net} (c) shows the Convolutional Block Attention Module (CBAM) composed of both channel attention and spatial attention. 
Sections \ref{CBAM} provide a detailed explanation of the algorithm's computational principles.

\begin{table}[tb]
	\caption{FU-CBAM-Net Encoder Block} \label{FU-CBAM_enecoder}
	\begin{center}
	 	\begin{tabular}{c|cc|c}
	 		\hline \hline
	 		& \multicolumn{2}{c|}{FU-CBAM-Net Encoder Block}                                                                                             & Feature size \\ \hline
	 		Input                         & \multicolumn{2}{c|}{-}                                                                                                       & $B \times 15 \times 61 \times 299$  \\ \hline
	 		Encoder layer 1               & \multicolumn{2}{c|}{\begin{tabular}[c]{@{}c@{}}
	 				
	 		$\left[\begin{array}{ccc}
	 			conv1, & 3 \times 3 \\
	 			conv2, & 3 \times 3 \\
	 			padding, & 1 
	 		\end{array}\right]$
 		\end{tabular}}                      &  $B \times 64 \times 61 \times 299$  \\ \hline
	 		\multirow{2}{*}{ \centering CBAM  layer 1} & \multicolumn{1}{c|}{Channel Attention}  & \begin{tabular}[c]{@{}c@{}}Adaptive average pool, $1 \times 1$ \\ Adaptive max pool, $1 \times 1$\end{tabular} & $B \times 64 \times 1 \times 1$     \\ \cline{2-4} 
	 		& \multicolumn{1}{c|}{Spatial Attention}  & \begin{tabular}[c]{@{}c@{}}
	 		$\left[\begin{array}{ccc}
	 		conv, & 7 \times 7 \\
	 		stride, & 1 \\
	 		padding, & 3 
	 		\end{array}\right]$
	 		\end{tabular}              & $B \times 1 \times 61 \times 299$   \\ \hline
	 		Encoder layer 2               & \multicolumn{2}{c|}{\begin{tabular}[c]{@{}c@{}}
	 		$\left[\begin{array}{ccc}
	 		conv1, & 3 \times 3 \\
	 		conv2, & 3 \times 3 \\
	 		max pool, & 2 \times 2 
	 		\end{array}\right]$
 	        \end{tabular}}                      & $B \times 128 \times 30 \times 149$ \\ \hline
	 		\multirow{2}{*}{CBAM layer 2} & \multicolumn{1}{c|}{Channel  Attention} & \begin{tabular}[c]{@{}c@{}}Adaptive average pool, $1 \times 1$ \\ Adaptive max pool, $1 \times 1$\end{tabular} & $B \times 128 \times 1 \times 1$    \\ \cline{2-4} 
	 		& \multicolumn{1}{c|}{Spatial  Attention} & \begin{tabular}[c]{@{}c@{}}
	 		$\left[\begin{array}{ccc}
	 		conv, & 7 \times 7 \\
	 		stride, & 1 \\
	 		padding, & 3 
	 		\end{array}\right]$
            \end{tabular}              & $B \times 1 \times 30 \times 149 $  \\ \hline
	 		Encoder layer 3               & \multicolumn{2}{c|}{\begin{tabular}[c]{@{}c@{}}
	 		$\left[\begin{array}{ccc}
	 		conv1, & 3 \times 3 \\
	 		conv2, & 3 \times 3 \\
	 		max pool, & 2 \times 2 
	 		\end{array}\right]$		
 			\end{tabular}}                      & $B \times 256 \times 15 \times 74$  \\ \hline
	 		\multirow{2}{*}{CBAM layer 3} & \multicolumn{1}{c|}{Channel  Attention} & \begin{tabular}[c]{@{}c@{}}Adaptive average pool, $1 \times 1$ \\ Adaptive max pool, $1 \times 1$\end{tabular} & $B \times 256 \times 1 \times 1$    \\ \cline{2-4} 
	 		& \multicolumn{1}{c|}{Spatial Attention}  & \begin{tabular}[c]{@{}c@{}}
	 		$\left[\begin{array}{ccc}
	 		conv, & 7 \times 7 \\
	 		stride, & 1 \\
	 		padding, & 3 
	 		\end{array}\right]$
 	        \end{tabular}              & $B \times 1 \times 15 \times 74$   \\ \hline
	 		Encoder layer 4               & \multicolumn{2}{c|}{\begin{tabular}[c]{@{}c@{}}
	 		$\left[\begin{array}{ccc}
	 		conv1, & 3 \times 3 \\
	 		conv2, & 3 \times 3 \\
	 		max pool, & 2 \times 2 
	 		\end{array}\right]$			
			\end{tabular}}                      & $B \times 512 \times 7 \times 37$   \\ \hline
	 		\multirow{2}{*}{CBAM layer 4} & \multicolumn{1}{c|}{Channel  Attention} & \begin{tabular}[c]{@{}c@{}}Adaptive average pool, $1 \times 1$ \\ Adaptive max pool, $1 \times 1$\end{tabular} & $B \times 512 \times 1 \times 1$    \\ \cline{2-4} 
	 		& \multicolumn{1}{c|}{Spatial Attention}  & \begin{tabular}[c]{@{}c@{}}
	 		$\left[\begin{array}{ccc}
	 		conv, & 7 \times 7 \\
	 		stride, & 1 \\
	 		padding, & 3 
	 		\end{array}\right]$	
 			\end{tabular}              & $B \times 1 \times 7 \times 37$     \\ \hline
	 		Encoder layer 5               & \multicolumn{2}{c|}{\begin{tabular}[c]{@{}c@{}}
	 		$\left[\begin{array}{ccc}
	 		conv1, & 3 \times 3 \\
	 		conv2, & 3 \times 3 \\
	 		max pool, & 2 \times 2 
	 		\end{array}\right]$			
 	       \end{tabular}}                      & $B \times 1024 \times 3 \times 18$  \\ \hline \hline
	 	\end{tabular}
	\end{center} 
	\vspace{-1.5em}
\end{table} 

\begin{table}[tb]
	\caption{FU-CBAM-Net Decoder Block} \label{FU-CBAM_decoder}
	\begin{center}
		\begin{tabular}{c|cc|c}
			\hline \hline
			& \multicolumn{2}{c|}{FU-CBAM-Net Decoder}                                  & Feature size                  \\ \hline
			Decoder layer 1                  & \multicolumn{2}{c|}{\begin{tabular}[c]{@{}c@{}}
			$\left[\begin{array}{ccc}
			conv, & 3 \times 3 \\
			stride, & 2 \\
			dilation, & 2 \\
			padding, & 1 \\
			output padding, &0
			\end{array}\right]$	
	        \end{tabular}}                       & $B \times
         512 \times 7 \times 37$                    \\ \hline
			\multirow{2}{*}{ \centering Decoder layer 2} & \multicolumn{1}{c|}{convolutionn layer} & \begin{tabular}[c]{@{}c@{}}
			$\left[\begin{array}{ccc}
			conv1, & 3 \times 3  \\
			conv2, & 3 \times 3
			\end{array}\right]$	
			\end{tabular}                                               & \multirow{2}{*}{ \centering $B \times 256 \times 15 \times 74$}  \\ \cline{2-3}
			& \multicolumn{1}{c|}{upsampling layer}   & \begin{tabular}[c]{@{}c@{}}
			$\left[\begin{array}{ccc}
			conv, & 4 \times 3  \\
			stride, & 2 \\
			padding, & 1 \\
			dilation, & 1 \\
			output padding, & 1
			\end{array}\right]$		
		
	        \end{tabular}   &                               \\ \hline
			\multirow{2}{*}{Decoder layer 3} & \multicolumn{1}{c|}{convolutionn layer} & \begin{tabular}[c]{@{}c@{}}
			$\left[\begin{array}{ccc}
		     conv1, & 3 \times 3 \\
		     conv2, & 3 \times 3
			\end{array}\right]$	
	        \end{tabular}                                               & \multirow{2}{*}{$B \times 128 \times 30 \times 147$} \\ \cline{2-3}
			& \multicolumn{1}{c|}{upsampling layer}   & \begin{tabular}[c]{@{}c@{}}
			$\left[\begin{array}{ccc}	
			conv, &3 \times 4\\ 
			stride, &2 \\ 
			padding, &1 \\ 
			dilation, &1\\ 
			output padding, &1
		    \end{array}\right]$	 
	        \end{tabular} &                               \\ \hline
			\multirow{2}{*}{Decoder layer 4} & \multicolumn{1}{c|}{convolutionn layer} & \begin{tabular}[c]{@{}c@{}}
			$\left[\begin{array}{cc}
			conv1, & 3 \times 3 \\ 
			conv2, & 3 \times 3
			\end{array}\right]$
			 \end{tabular}                                               & \multirow{2}{*}{$B \times 64 \times 61 \times 299$}  \\ \cline{2-3}
			& \multicolumn{1}{c|}{upsampling layer}   & \begin{tabular}[c]{@{}c@{}}
				$\left[\begin{array}{cc}
				conv, &4 \times 4\\ 
				stride, &2 \\ 
				padding, &1
			\end{array}\right]$	
			 \end{tabular}                                 &                               \\ \hline
			Decoder layer 5                  & \multicolumn{2}{c|}{\begin{tabular}[c]{@{}c@{}}
			$\left[\begin{array}{cc}		
			 conv1, & 3 \times 3\\ 
			 conv2, & 3 \times 3 \\ 
			 conv3, & 1 \times 1
			\end{array}\right]$	
		    \end{tabular}}                                                       & $B \times 9 \times 61 \times 299$                    \\ \hline \hline
		\end{tabular}
	\end{center} 
	\vspace{-1.5em}
\end{table} 

\subsubsection{Convolutional Block Attention Module} \label{CBAM}

\begin{figure*}[!h]
	\begin{center}
		\includegraphics[width=1 \linewidth]{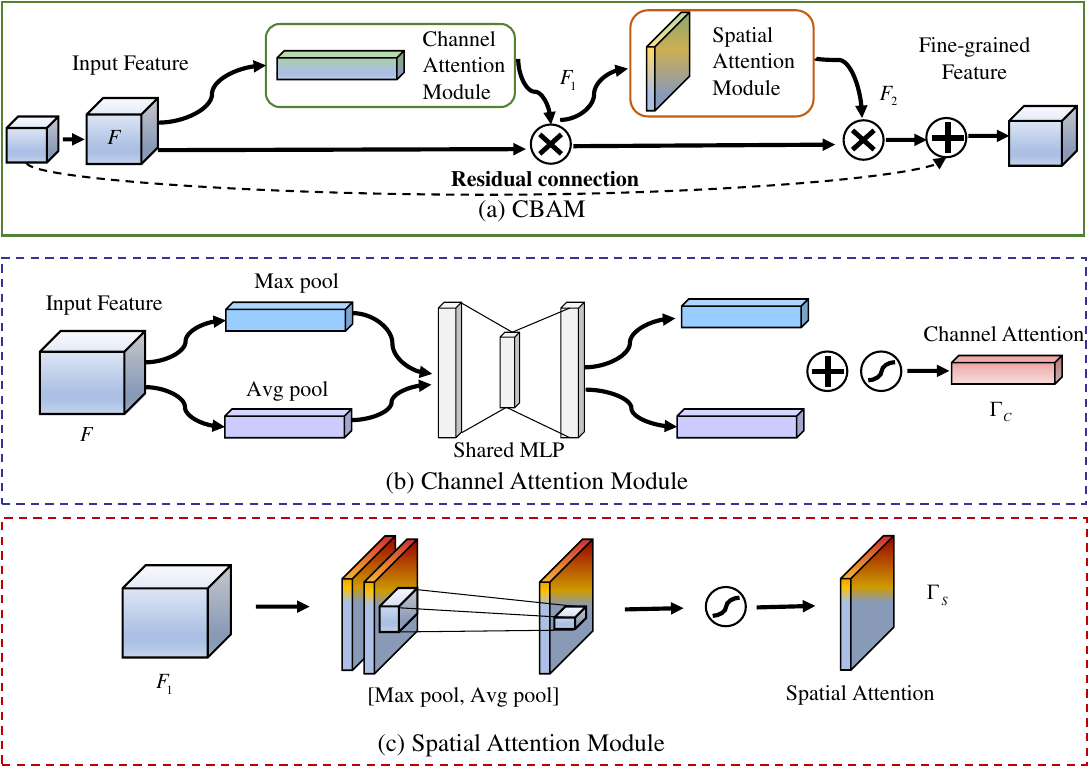}
	\end{center}  \vspace{-2mm} 
	\caption{{{Convolutional Block Attention Module network architecture.}
	}} \label{CBAM_1} 
\end{figure*}

Figure \ref{CBAM_1} presents the detailed network architecture diagram of CBAM. 
For the input feature map $F \in \mathbb{R}^{C \times H \times W}$, after passing through the channel attention layer and the spatial attention layer, the feature map sizes are $\Gamma_c \in \mathbb{R}^{C \times 1 \times 1} $ and $\Gamma_s \in \mathbb{R}^{C\times H \times W}$, respectively.
The overall computation process of CBAM is as follows:
\begin{equation} \label{cbam_equation}
  \left\{
  	\begin{aligned}
  		F_1 &= \Gamma_c(F) \otimes F  \\
  		F_2 &= \Gamma_s(F_1) \otimes F_1 \\
  		\Gamma_c(F) &=  \sigma(MLP(AvgPoll(F)) + MLP(MaxPool(F))) \\
  		\Gamma_s(F_1) &= \sigma(f^{7 \times 7}([AvgPool(F_1); MaxPool(F_1)]))
  	\end{aligned}
  \right.	
\end{equation}

In the above Eq. \eqref{cbam_equation}, the symbol $\otimes$ represents element-wise multiplication. 
$f^{7 \times 7}$ denotes a convolutional operation with a $7 \times 7$ filter size.
$\sigma$ denotes the sigmoid fuction.
This function is often used as the activation function in neural networks to map variables to the range [0, 1]. The calculation formula is as follows:

\begin{equation}
	S(x) = \frac{1}{1+e^{-x}}
\end{equation}

In Fig. \ref{CBAM_1} (a), the feature map $F$, input to a layer in the network, undergoes channel attention to produce the output $\Gamma_c(F)$, which is then element-wise multiplied with itself to obtain $F_1$.
Following that, $F_1$ is fed into the spatial attention layer to obtain the output $\Gamma_s(F_1)$, which is then element-wise multiplied with $F_1$ to produce the output $F_2$.
Finally, $F_2$ is combined with the input feature map $F$ through a residual connection to obtain a fine-grained output feature $\varrho$.

Fig. \ref{CBAM_1} (b) illustrates the channel attention module in CBAM.
For computational efficiency, the input feature maps are compressed into one dimension, and different features in space are aggregated using both max-pooling and average-pooling operations.
Subsequently, the aggregated features are each processed through a shared MLP layer, and the final output channel attention feature map $\Gamma_c \in \mathbb{R}^{C \times 1 \times 1}$ is obtained by element-wise summation.

The spatial attention module in CBAM is depicted in Fig. \ref{CBAM_1} (c).
First, perform average-pooling and max-pooling operations along the channel dimension.
Then, aggregate the extracted features using a convolutional neural network with a $7 \times 7$ kernel to obtain a two-dimensional spatial attention map $\Gamma_s \in \mathbb{R}^{1 \times H \times W}$.

\section{Data preparation} \label{section 3}

\begin{figure*}[!h]
	\begin{center}
		\includegraphics[width=0.5 \linewidth]{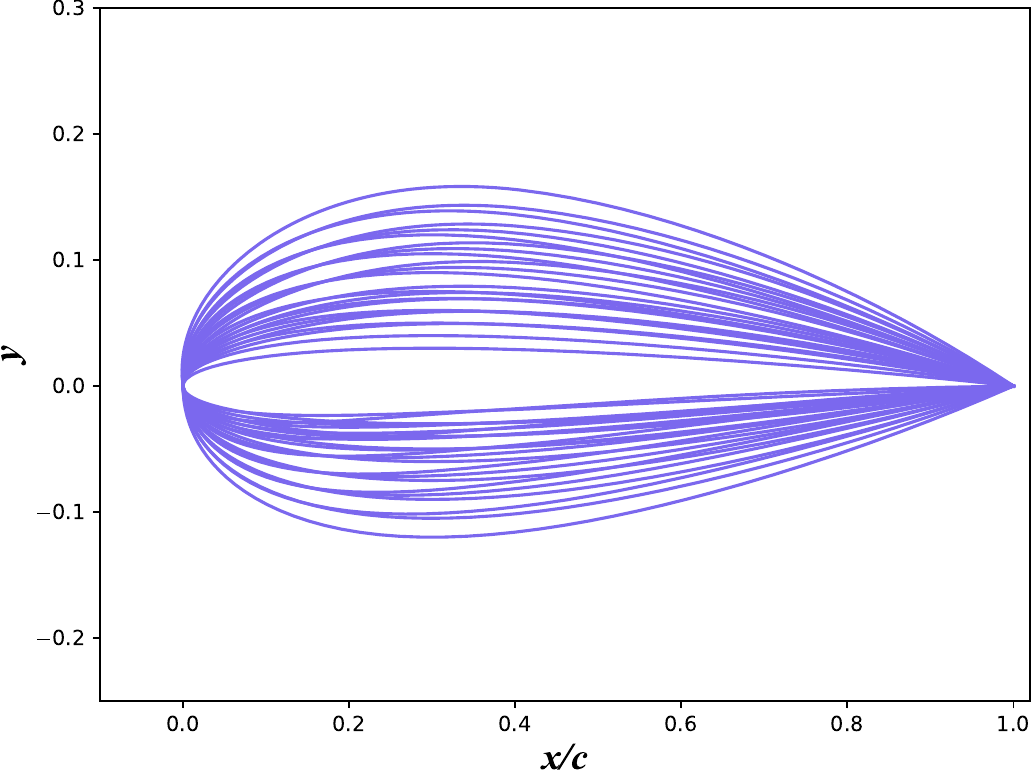}
	\end{center}  \vspace{-2mm}  
	\caption{{{NACA airfoil dataset.}
	}} \label{naca airfoils} 
\end{figure*}

Utilize the FU-CBAM-Net neural network introduced in Section \ref{fu-cbam-net} to accomplish the flow field prediction task for different airfoils under various operating conditions.
Test data for 24 NACA-series airfoils is depicted in Fig. \ref{naca airfoils}.
Using the PHengLEI solver to calculate the laminar flow solutions for these 24 airfoils within the Reynolds number range of 1000 to 2000 (1000, 1200, 1400, 1600, 1800, 2000) and an angle of attack ranging from 0\degree\ to 5\degree (0\degree, 1\degree, 2\degree, 3\degree, 4\degree, 5\degree).
A total of 864 flow field datasets are generated for training and testing tasks in the FU-CBAM-Net neural network.
$80\%$ of the data will be used as the training dataset, $10\%$ as the cross-validation dataset, and the remaining $10\%$ as the testing dataset.

For a single case, there are a total of 18239 structured grid flow field data points. 
The input for FU-CBAM-Net consists of parameters in 15 channels ($x$, $y$, $x_0$, $y_0$, $\xi$, $\eta$, $SDF$, $M_x$, $M_y$, $AOA$, $Re$, $\frac{\partial x}{\partial \xi}$, $\frac{\partial x}{\partial \eta}$, $\frac{\partial y}{\partial \xi}$, $\frac{\partial y}{\partial \eta}$), and the output consists of parameters in 9 channels ($u$, $v$, $Cp$, $\frac{\partial u}{\partial x}$, $\frac{\partial u}{\partial y}$, $\frac{\partial v}{\partial x}$, $\frac{\partial u}{\partial y}$, $\frac{\partial Cp}{\partial x}$,
$\frac{\partial Cp}{\partial y}$). 
Therefore, for FU-CBAM-Net, the neural network input parameter size is $15 \times 61 \times 299$, and the output parameter size is $9 \times 61 \times 299$.

\section{Results and discussions}

\subsection{Discussion of training results}

Based on the flow field data introduced in Section \ref{section 3}, training is conducted for UNet, UNet3+, and the FU-CBAM-Net, separately.
The initial learning rate for the neural network model during the training process is set to $5 \times 10^{-5}$, and the Adam optimizer is employed to train the neural network parameters.
The model is implemented using the PyTorch deep learning library in the Python programming language. 
We performed the training tasks on an NVIDIA RTX 3090 GPU under the Linux platform.
The mean squared error (MSE) is used as the loss function for the model during the training process, and its calculation formula is as follows:

\begin{equation} \label{eq13}
\begin{aligned}
	MSE_{loss} = \frac{1}{9 \times N} \sum_{i=1}^{N}[(u_i^t - u_i^p)^2 + (v_i^t - v_i^p)^2 + (Cp_i^t - Cp_i^p)^2 + \\
	 ((\frac{\partial u}{\partial \xi})_i^t - (\frac{\partial u}{\partial \xi})_i^p)^2 +  
	 ((\frac{\partial u}{\partial \eta})_i^t - (\frac{\partial u}{\partial \eta})_i^p)^2+
	 ((\frac{\partial v}{\partial \xi})_i^t - (\frac{\partial v}{\partial \xi})_i^p)^2 +  \\
	  ((\frac{\partial v}{\partial \eta})_i^t - (\frac{\partial v}{\partial \eta})_i^p)^2 + 
	   ((\frac{\partial Cp}{\partial \xi})_i^t - (\frac{\partial Cp}{\partial \xi})_i^p)^2 +
	    ((\frac{\partial Cp}{\partial \eta})_i^t - (\frac{\partial Cp}{\partial \eta})_i^p)^2]
\end{aligned}
\end{equation}

In the above Eq. \eqref{eq13}, $\wp_i^t$ and $\wp_i^p$ ($\wp$: $u , v, Cp, \frac{\partial u}{\partial \xi}, \frac{\partial u}{\partial \eta}, \frac{\partial v}{\partial \xi}, \frac{\partial v}{\partial \eta}, \frac{\partial Cp}{\partial \xi}, \frac{\partial Cp}{\partial \eta}$) respectively represent the ground-truth values and the neural network predicted values.

\begin{figure*}[!h]
	\begin{center}
		\includegraphics[width=1 \linewidth]{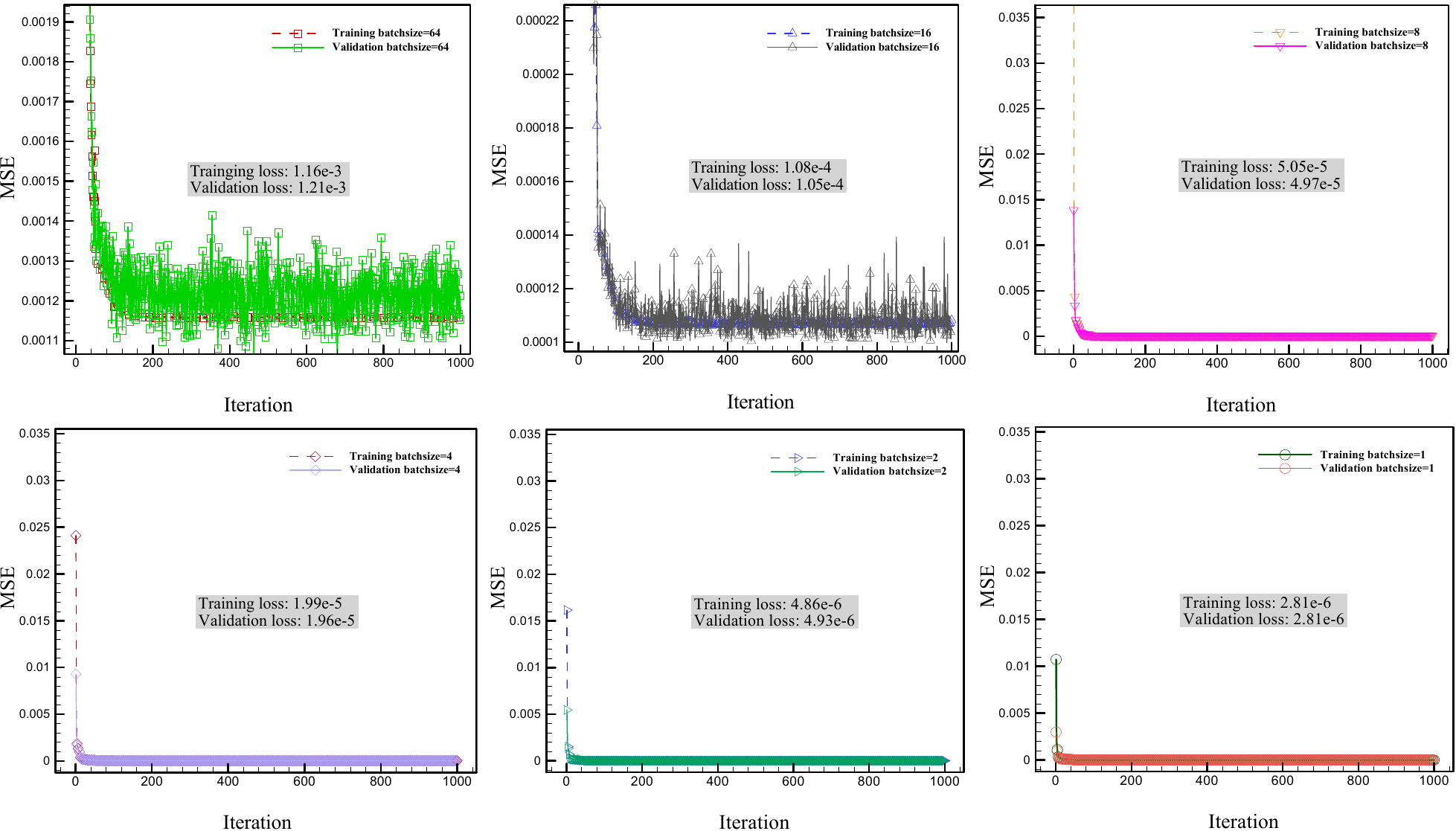}
	\end{center}  \vspace{-2mm}  
	\caption{{{The variation curve of the model training loss function for the UNet neural network under different batchsize.}
	}} \label{unet loss} 
\end{figure*}

As shown in Fig. \ref{unet loss}, for the UNet neural network, we conducted ablation experiments on the parameter 'batchsize' to explore its impact on the training results of the neural network.
When the batchsize is set to 64, the curve of the MSE loss function on the cross-validation set exhibits pronounced oscillations. Additionally, the model's loss function remains relatively high even during convergence, with a training set loss of $1.16 \times 10^{-3}$ and a cross-validation set MSE loss of $1.21 \times 10^{-3}$.
As the batchsize decreases, the loss function curves on both the training and testing sets tend to stabilize. 
When the batchsize is set to 1, the loss function values reach the minimum, with an MSE loss of $2.81 \times 10^{-6}$ on both the training and testing sets.

\begin{figure*}[!h]
	\begin{center}
		\includegraphics[width=0.7 \linewidth]{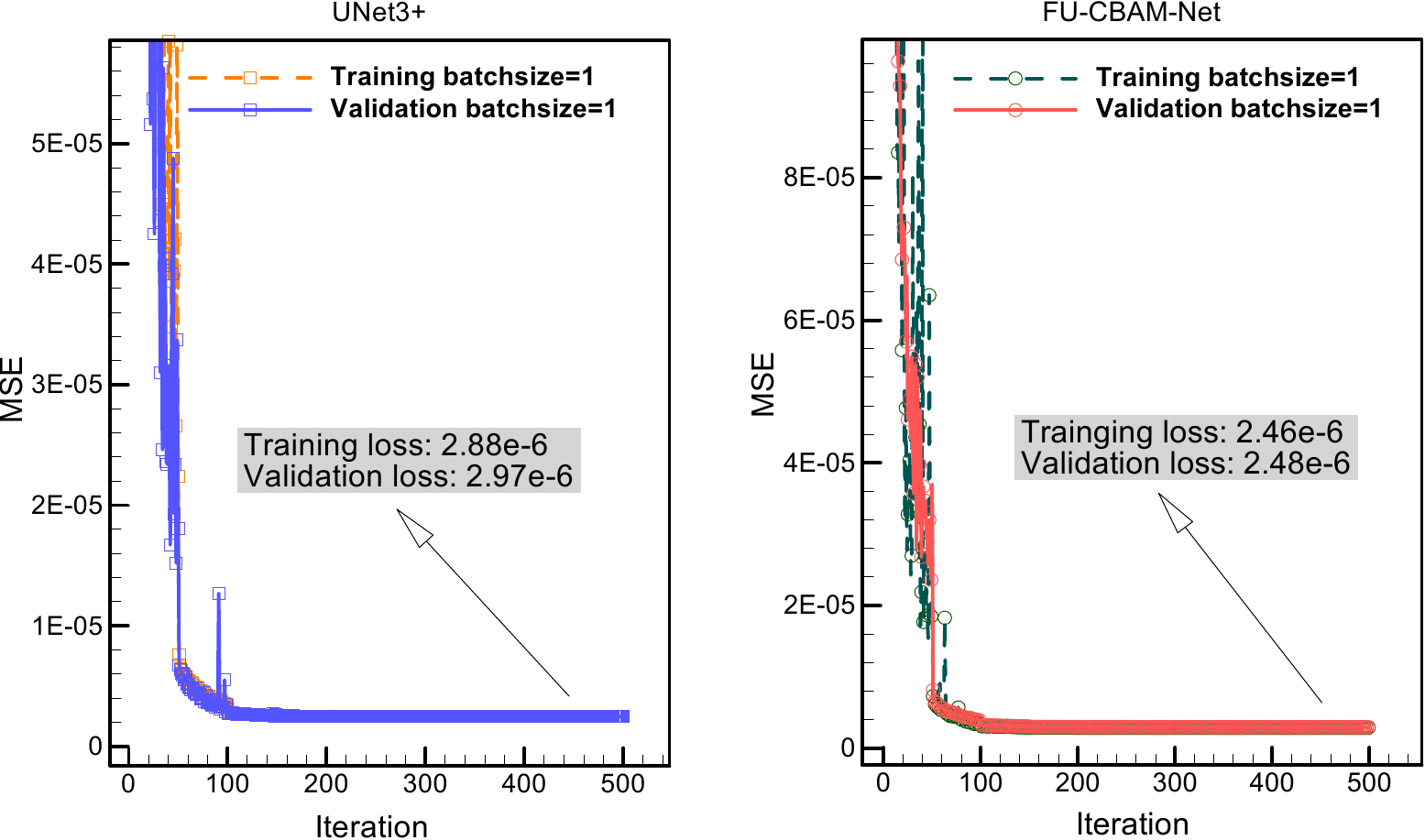}
	\end{center}  \vspace{-2mm}  
	\caption{{{The loss function variation curves during the training process for UNet3+ and FU-CBAM-Net (left: UNet3+, right: FU-CBAM-Net).}
	}} \label{fu-cbam loss} 
\end{figure*}

Figure \ref{fu-cbam loss} illustrates the loss function variation curves during the training process for the UNet3+ and FU-CBAM-Net neural network models. 
The hyperparameter batchsize for both models is set to 1 during the training process.
For UNet3+, the loss function curve converges rapidly before the epoch of 100.
After approximately 200 iterations, the curve stabilizes. 
The loss function for UNet3+ on the training set is $2.88 \times 10^{-6}$, and on the cross-validation set, it is $2.97 \times 10^{-6}$.
Similarly, for the proposed FU-CBAM-Net method, the curve descends rapidly before the 100th iteration. 
The incorporation of attention layers allows this method to achieve a smaller loss compared to UNet and UNet3+, indicating higher predictive accuracy for the deep learning model.

\subsection{Flow field prediction} \label{flow field prediction results}

\begin{figure*}[!h]
	\begin{center}
		\includegraphics[width=0.9 \linewidth]{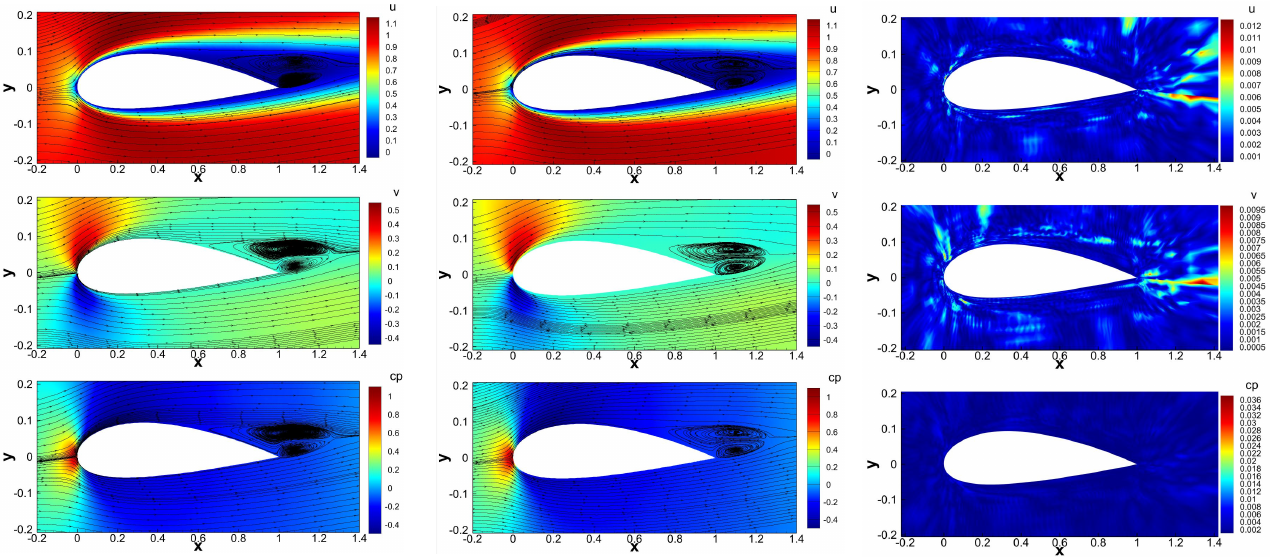}
	\end{center}  \vspace{-2mm}  
	\caption{{{Comparison between the PHengLEI computational results and the predicted results of the FU-CBAM-Net for NACA2415 Re=1600, AOA=3\degree (left: PHengLEI calculation results, middle: FU-CBAM-Net prediction results, right: absolute error map between PHengLEI and FU-CBAM-Net).}
	}} \label{naca2415_1600_3} 
\end{figure*}

\begin{figure*}[!h]
	\begin{center}
		\includegraphics[width=0.9 \linewidth]{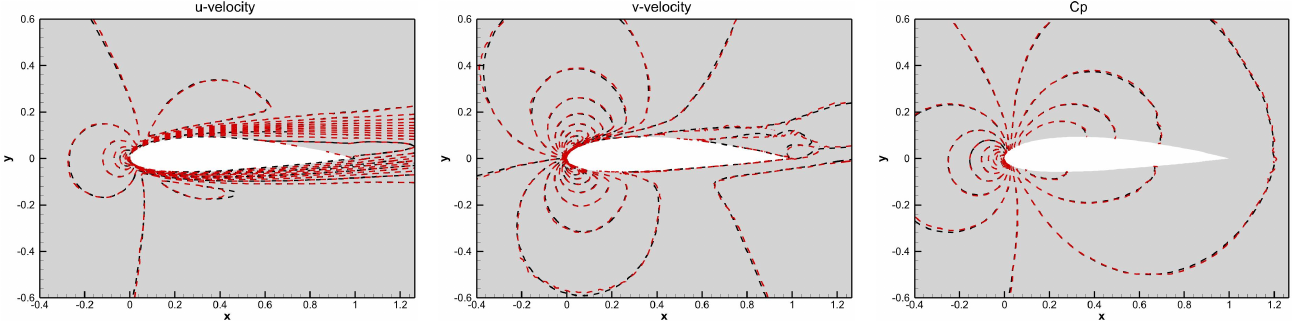}
	\end{center}  \vspace{-2mm}  
	\caption{{{Contour comparison diagram of PHengLEI and FU-CBAM-Net with respect to NACA2415 Re=1600 AOA=3\degree. The black dashed lines represent the calculation results by PHengLEI, while the red dashed lines represent the predictive results from FU-CBAM-Net.}
	}} \label{naca2415_1600_3_contour} 
\end{figure*}

In this section, we assess the accuracy and generalization of the prediction results of the FU-CBAM-Net neural network using data from a testing dataset that wasn't utilized during the training process of the model.
Firstly, we tested the trained neural network model, FU-CBAM-Net, for its predictive performance on the flow around the NACA2415 airfoil at Re=1600 and AOA=3 \degree.
As shown in Fig. \ref{naca2415_1600_3}, the predictive results of FU-CBAM-Net closely match the computed results of PHengLEI.
Further analysis from the absolute error plots in the last column of Fig. \ref{naca2415_1600_3} reveals that for u-velocity, the absolute error ranges from $1 \times 10^{-3}$ to $1.2 \times 10^{-2}$; for v-velocity, the error spans from $5 \times 10^{-4}$ to $9.5 \times 10 ^{-3}$. Meanwhile, for pressure coefficient Cp, the error range lies between $2 \times 10^{-3}$ and $3.6 \times 10^{-2}$.
Figure \ref{naca2415_1600_3_contour} also provides contour diagrams comparing the predicted results from the neural network with the computational results from PHengLEI. 
Whether for u-velocity, v-velocity, or Cp, the curves of the predicted results align closely with the computational results from PHengLEI, demonstrating that the neural network FU-CBAM-Net has achieved a high level of predictive accuracy.

\begin{figure*}[htbp]
	\subfloat[u-velocity]{\includegraphics[width=0.3\textwidth]{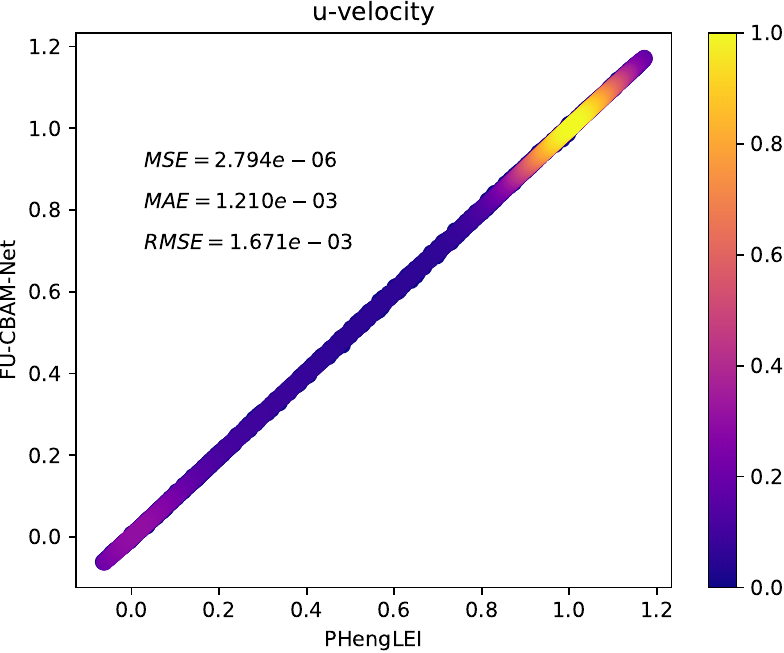}}
	\hfill
	\subfloat[v-velocity]{\includegraphics[width=0.3\textwidth]{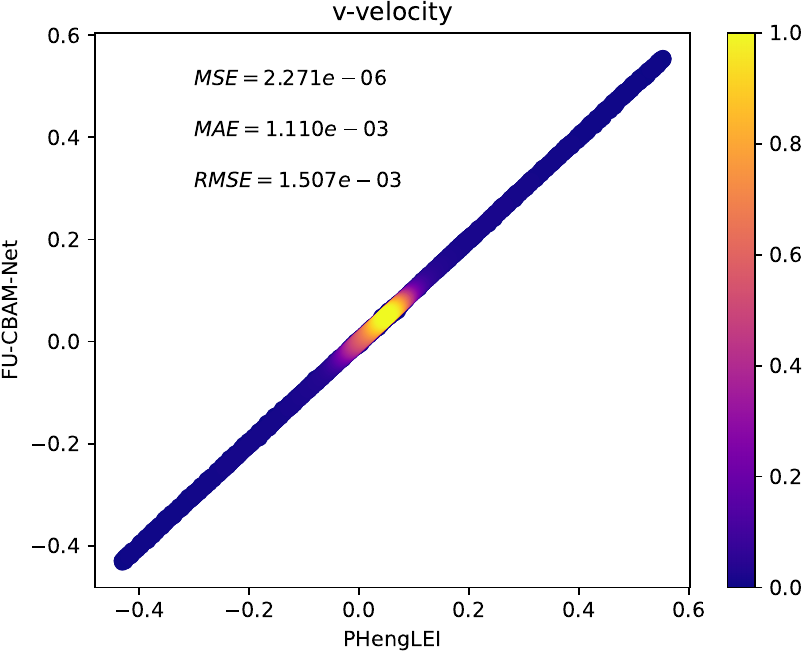}}
	\hfill
	\subfloat[Cp]{\includegraphics[width=0.3\textwidth]{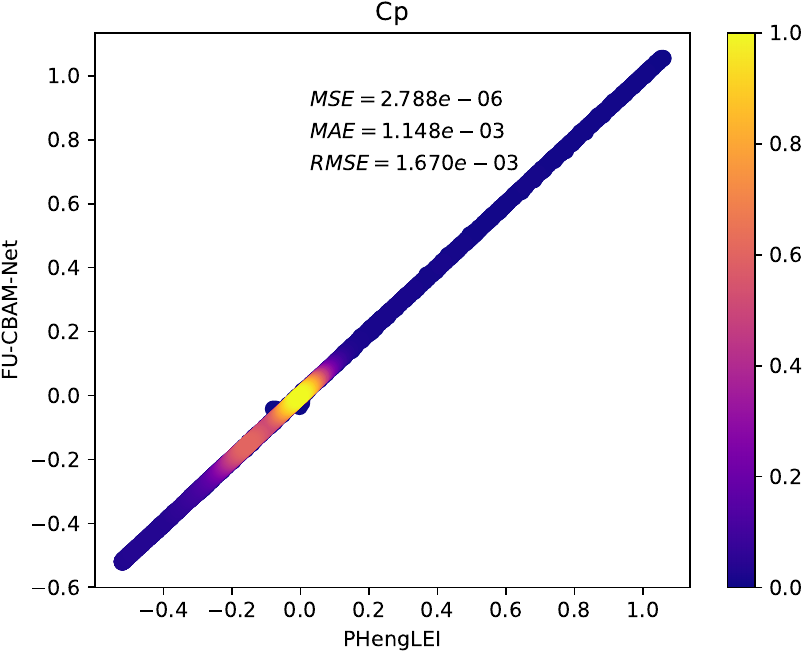}}
	\caption{Scatter density diagram between PHengLEI and FU-CBAM-Net for NACA2415 Re=1600 AOA=3\degree.}
	\label{NACA2415 1600 3 scatter density}
\end{figure*}

To further examine the predictive accuracy of FU-CBAM-Net, Fig. \ref{NACA2415 1600 3 scatter density} depicts the scatter density plot between the PHengLEI and the FU-CBAM-Net.
The diagram also presents the mean squared error (MSE), mean absolute error (MAE), and root mean square error (RMSE) between the predicted values and the ground-truth.
Here, the predicted results of the airfoil flow field are defined as $\hat{\vartheta}=\{\hat{\vartheta}_1, \hat{\vartheta}_2,...,\hat{\vartheta}_n\}$, and the ground-truth of the flow field are defined as $\vartheta=\{\vartheta_1, \vartheta_2,...,\vartheta_n\}$.
MSE, MAE, and RMSE can be defined by the following formulas:
\begin{equation}
	\begin{aligned}
 	  MSE &= \frac{1}{n}\sum_{i=1}^{n}(\hat{\vartheta}_i - \vartheta_i)^2 \\
	  MAE &= \frac{1}{n}\sum_{i=1}^{n}|\hat{\vartheta}_i - \vartheta_i| \\
	  RMSE &= \sqrt{\frac{1}{n}\sum_{i=1}^{n}(\hat{\vartheta}_i - \vartheta_i)^2}
	\end{aligned}
\end{equation}

In Fig. \ref{NACA2415 1600 3 scatter density}, the scatter density plot between the calculation results of PHengLEI and the predicted results of the neural network is almost a straight line with a diagonal distribution. 
In Fig. \ref{NACA2415 1600 3 scatter density}(a), the values of MSE, MAE, and RMSE are $2.794 \times 10^{-6}$, $1.210 \times 10^{-3}$, and $1.671 \times 10^{-3}$, respectively.
For the v-velocity in Fig. \ref{NACA2415 1600 3 scatter density}(b), the values of MSE, MAE, and RMSE are  $2.271 \times 10^{-6}$,  $1.110 \times 10^{-3}$, and  $1.507 \times 10^{-3}$, respectively.
For the Cp in Fig. \ref{NACA2415 1600 3 scatter density}(c), the values of MSE, MAE, and RMSE are  $2.788 \times 10^{-6}$,  $1.148 \times 10^{-3}$, and  $1.670 \times 10^{-3}$, respectively.
By comparing different indicators, the accuracy of the flow field prediction results of the FU-CBAM-Net neural network has been further verified.

\begin{figure*}[htbp]
	\centering
	\subfloat[u-velocity]{\includegraphics[width=0.3\textwidth]{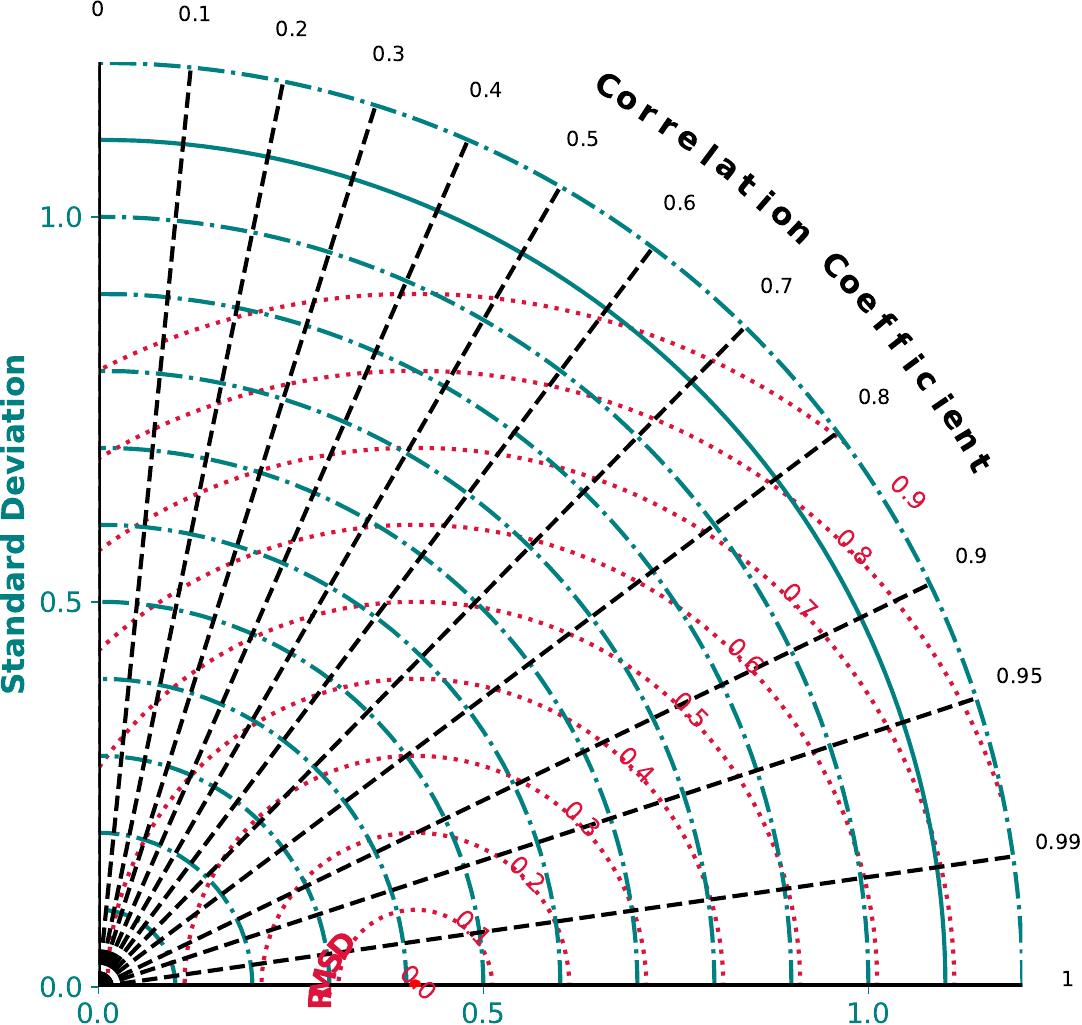}}
	\hfill
	\subfloat[v-velocity]{\includegraphics[width=0.3\textwidth]{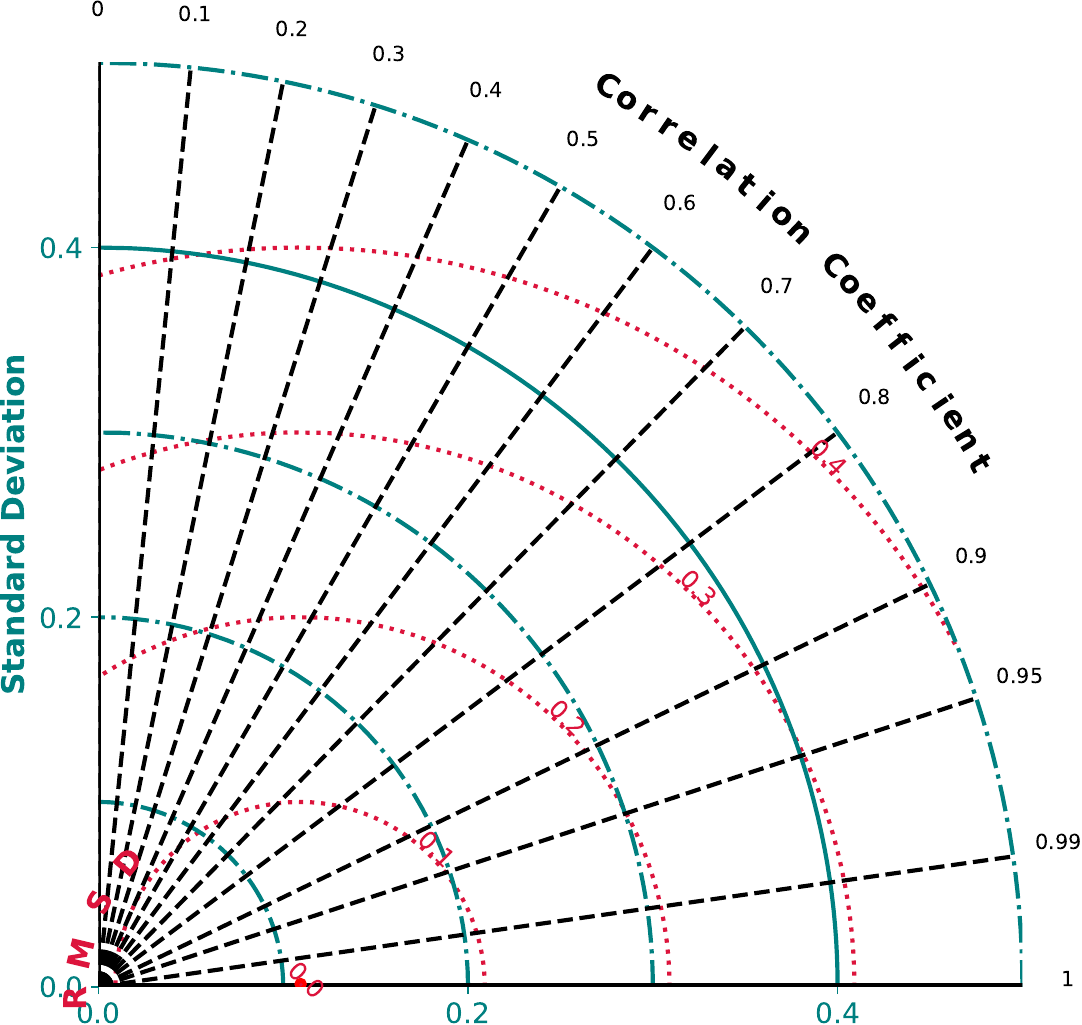}}
	\hfill
	\subfloat[Cp]{\includegraphics[width=0.3\textwidth]{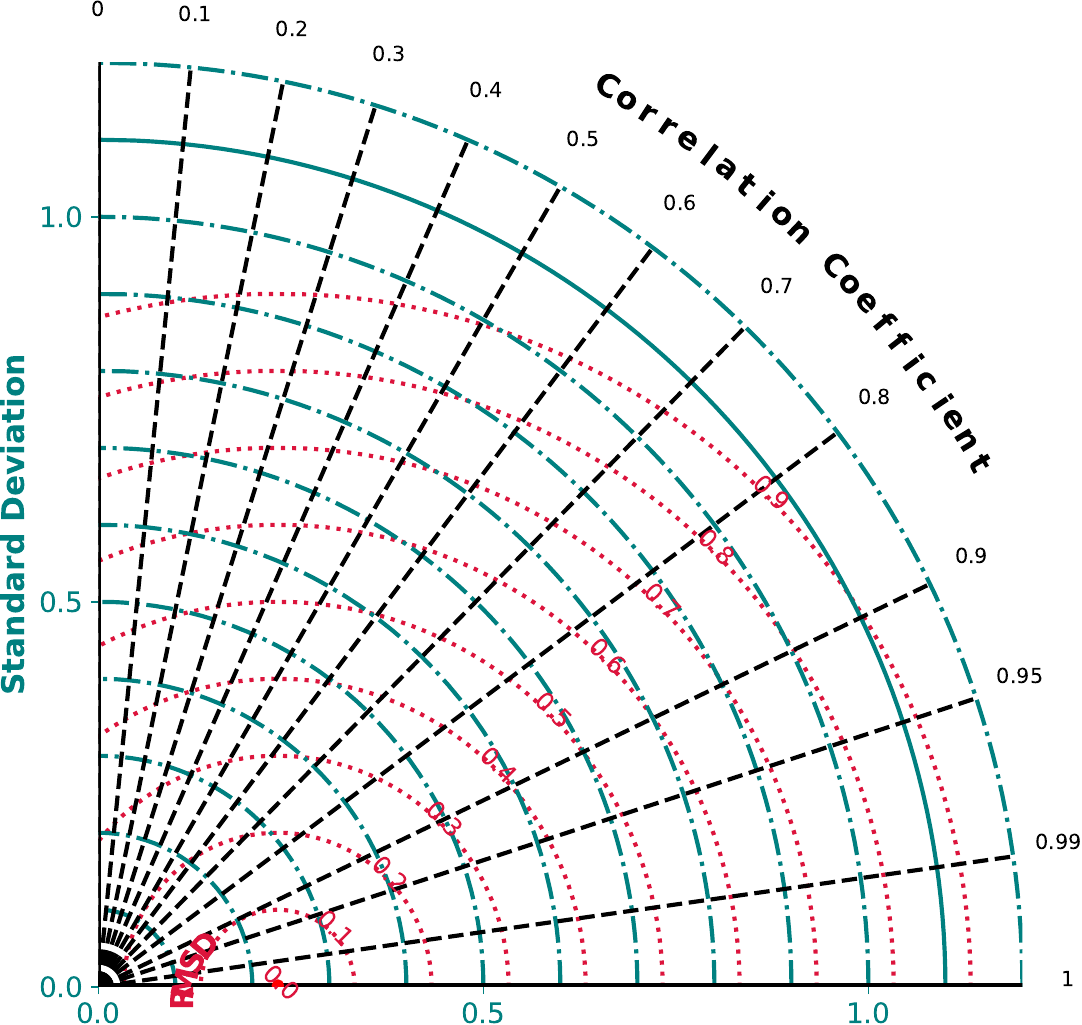}}
	\caption{Taylor diagram between PHengLEI and FU-CBAM-Net for NACA2415 Re=1600 AOA=3\degree.}
	\label{Taylor diagram naca2415 1600 3}
\end{figure*}

To further test the prediction effect of FU-CBAM-Net, Fig. \ref{Taylor diagram naca2415 1600 3} shows the Taylor diagram between the FU-CBAM-Net and PHengLEI.
The Taylor diagram is a visual representation that simultaneously displays three metrics: correlation coefficients, centered root-mean-square error (CRMSE), and standard deviation.
In Fig. \ref{Taylor diagram naca2415 1600 3}, the horizontal and vertical axes represent standard deviation, while the black radial lines depict correlation coefficients, and the red dashed lines represent CRMSE.

The correlation coefficients for u-velocity, v-velocity, and Cp are highly close to 1. 
Additionally, the CRMSE tends toward 0.
The quantitative analysis results further indicate that the predictive accuracy of FU-CBAM-Net is sufficiently high.
Please refer to \ref{Taylor diagram} for a more detailed description of the Taylor diagram.

\begin{figure*}[!h]
	\begin{center}
		\includegraphics[width=1 \linewidth]{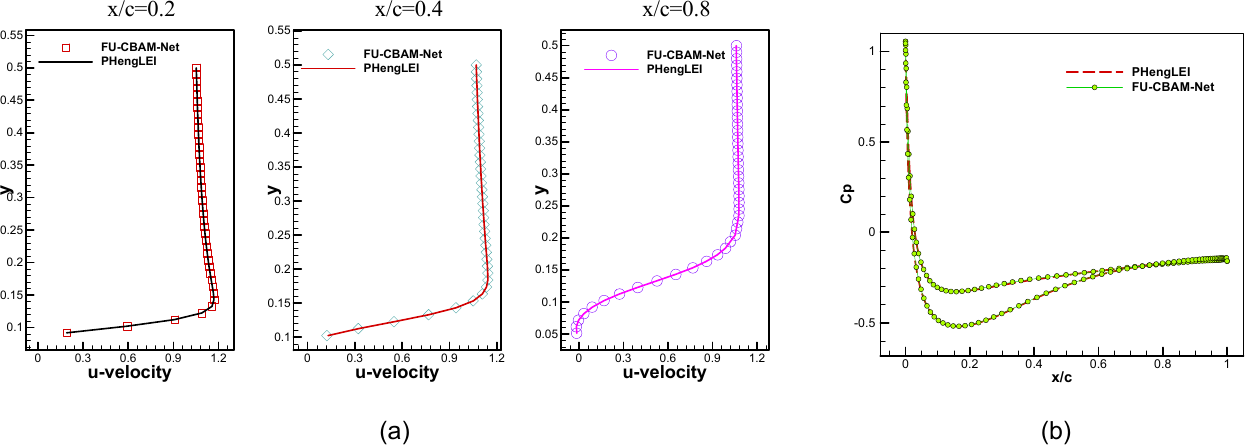}
	\end{center}  \vspace{-2mm}  
	\caption{{{Comparison between various velocity profiles and pressure coefficients for the reconstructed flow field and the ground-truth data. (a) velocity profiles. (b) Cp.}
	}} \label{naca2415_1600_3_cp} 
\end{figure*}

To test the predictive accuracy of the FU-CBAM-Net neural network for airfoil flow fields from various perspectives, Fig. \ref{naca2415_1600_3_cp} presents comparative curves between reconstructed velocity profiles at different locations and their ground-truth data, along with fitted curves for pressure coefficient Cp. 
From the comparative curves in Fig. \ref{naca2415_1600_3_cp}, it's evident that the velocity profiles from FU-CBAM-Net at different locations align perfectly with the computed results from PHengLEI.
Correspondingly, the pressure coefficient Cp also demonstrates a favorable predictive performance.

\begin{figure*}[!h]
	\begin{center}
		\includegraphics[width=0.9 \linewidth]{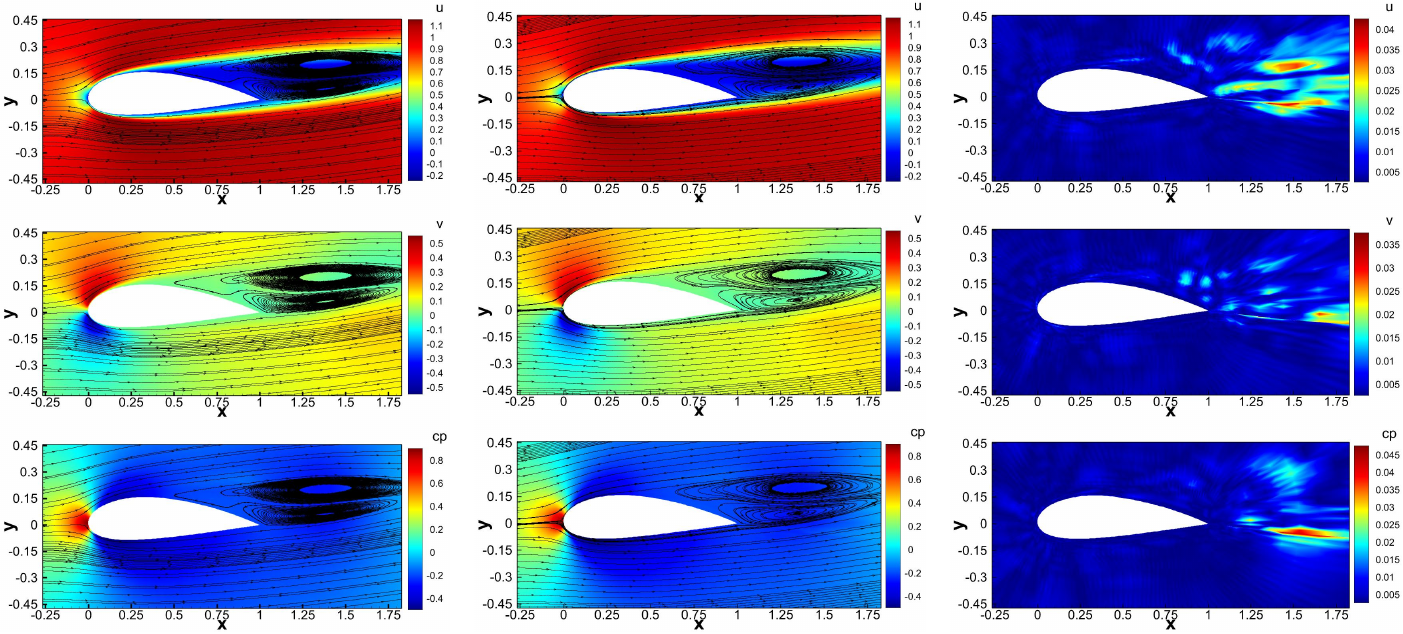}
	\end{center}  \vspace{-2mm}  
	\caption{{{Comparison between the PHengLEI computational results and the FU-CBAM-Net predicted results for NACA4424 Re=2000, AOA=5\degree (left: PHengLEI calculation results, middle: FU-CBAM-Net prediction results, right: absolute error map between PHengLEI and FU-CBAM-Net).}
	}} \label{naca4424_2000_5} 
\end{figure*}

\begin{figure*}[!h]
	\begin{center}
		\includegraphics[width=0.9 \linewidth]{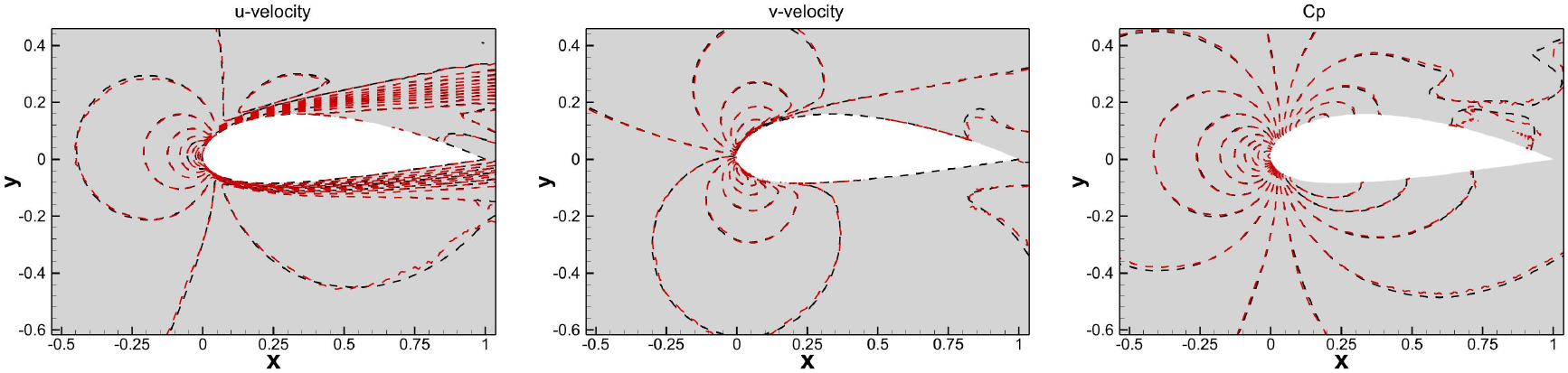}
	\end{center}  \vspace{-2mm}  
	\caption{{{Contour map between predicted and ground-truth values for NACA4424 Re=2000 AOA=5 \degree. The black dashed line represents the ground-truth values, while the red dashed line represents the predicted values.}
	}} \label{naca4424_2000_5_contour} 
\end{figure*}

Figure \ref{naca4424_2000_5} further illustrates the predicted flow field results of the FU-CBAM-Net neural network model with variations in airfoil geometry, Re, and AOA.
The predictive results reveal that even in the presence of strong separated flows, the neural network model still accurately simulates the flow field.
As observed from the absolute error plot on the right side of Fig. \ref{naca4424_2000_5}.
For u-velocity, the absolute error ranges from $5.0 \times 10^{-3}$ to $4.0 \times 10^{-2}$. 
For v-velocity, it ranges from $5.0 \times 10^{-3}$ to $3.5 \times 10^{-2}$. 
As for the pressure coefficient Cp, its absolute error ranges from $5 \times 10^{-3}$ to $4.5 \times 10^{-2}$.
Figure \ref{naca4424_2000_5_contour} further illustrates contour plots between the FU-CBAM-Net prediction results and PHengLEI calculation values. 
It's evident that for both velocity components, $u$ and $v$, as well as pressure coefficient Cp, the curves from both FU-CBAM-Net and PHengLEI align exceptionally well.

\begin{figure*}[htbp]
	\subfloat[u-velocity]{\includegraphics[width=0.3\textwidth]{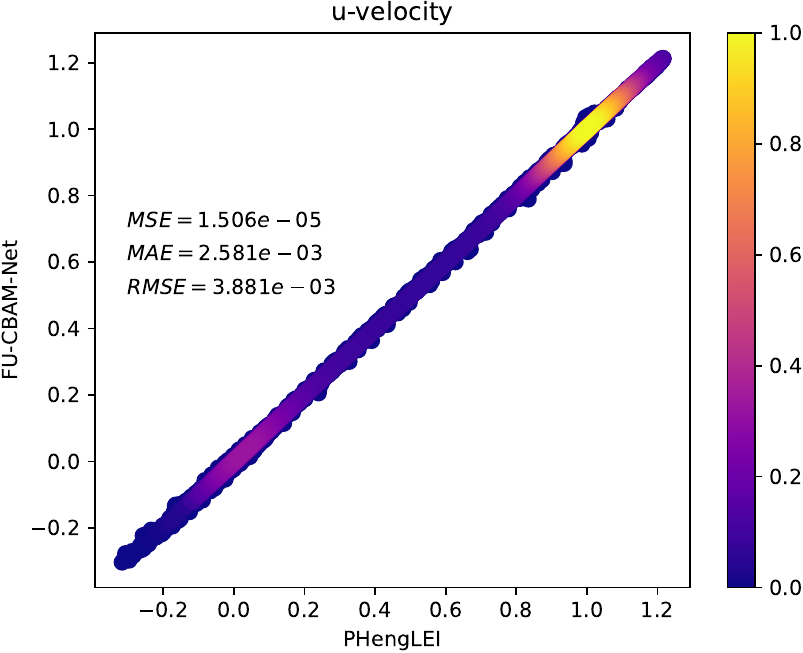}}
	\hfill
	\subfloat[v-velocity]{\includegraphics[width=0.3\textwidth]{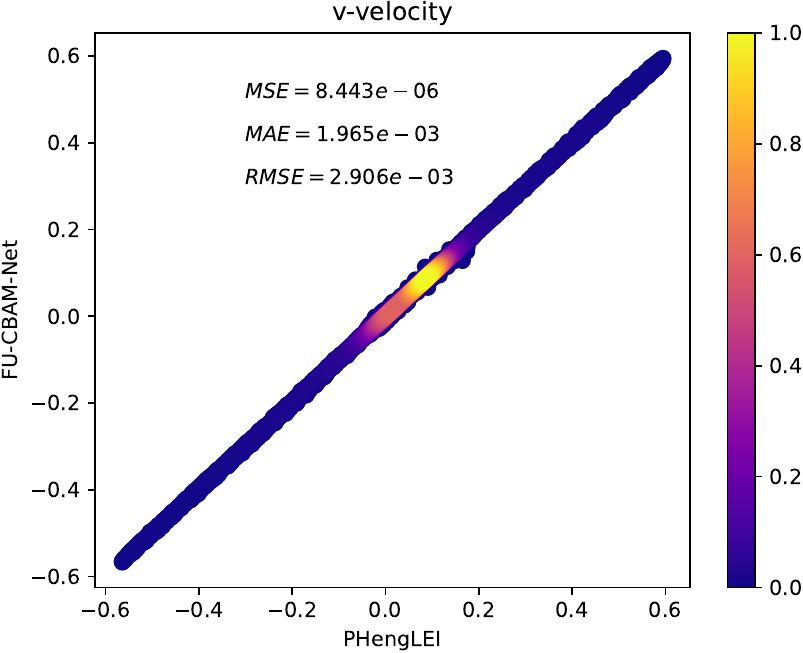}}
	\hfill
	\subfloat[Cp]{\includegraphics[width=0.3\textwidth]{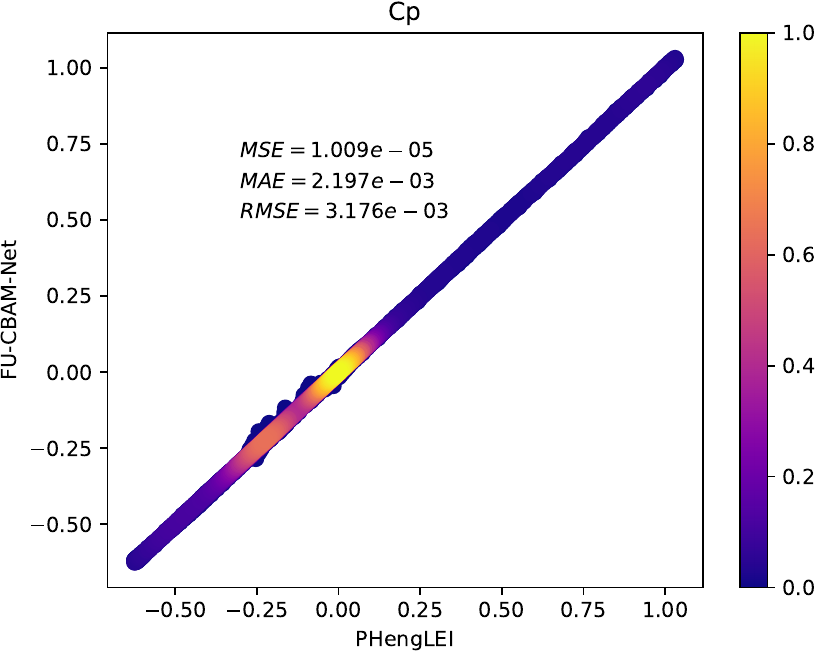}}
	\caption{Scatter density diagram between PHengLEI and FU-CBAM-Net for NACA4424 Re=2000 AOA=5\degree.}
	\label{NACA4424 2000 5 scatter density}
\end{figure*}

Figure \ref{NACA4424 2000 5 scatter density} presents scatter density diagrams for u-velocity, v-velocity, and pressure coefficient Cp. 
For u-velocity, v-velocity and Cp, the predicted and ground-truth values exhibit a diagonal distribution, indicating a strong correlation between them. 
Furthermore, provide the MSE, MAE, and RMSE values between the predicted results of FU-CBAM-Net and the computed results by PHengLEI.
In Fig. \ref{NACA4424 2000 5 scatter density}(a), the u-velocity takes the values $MSE=1.506 \times 10^{-5} $, $MAE=2.581 \times 10^{-3}$, and $RMSE=3.881 \times 10^{-3}$.
For v-velocity in Fig. \ref{NACA4424 2000 5 scatter density}(b), the numerical values are $MSE=8.443 \times 10^{-6}$, $MAE=1.965 \times 10^{-3}$, and $RMSE=2.906 \times 10^{-3}$.
Pressure coefficient Cp corresponds to the values $MSE=1.009 \times 10^{-5}$, $MAE=2.197 \times 10^{-3}$, and $RMSE=3.176 \times 10^{-3}$.
Above results further emphasizes that FU-CBAM-Net not only achieves good predictive accuracy but also demonstrates excellent generalization.

\begin{figure*}[htbp]
	\centering
	\subfloat[u-velocity]{\includegraphics[width=0.3\textwidth]{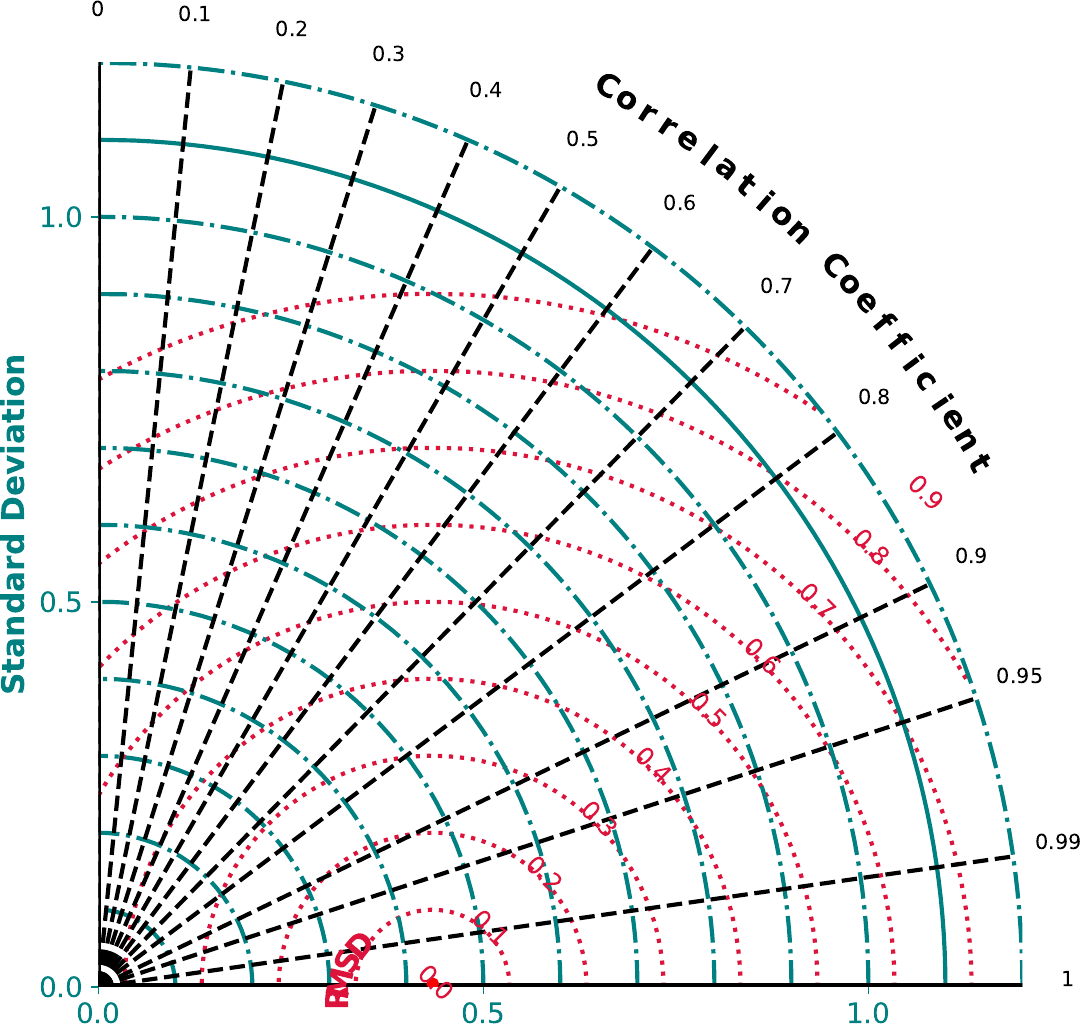}}
	\hfill
	\subfloat[v-velocity]{\includegraphics[width=0.3\textwidth]{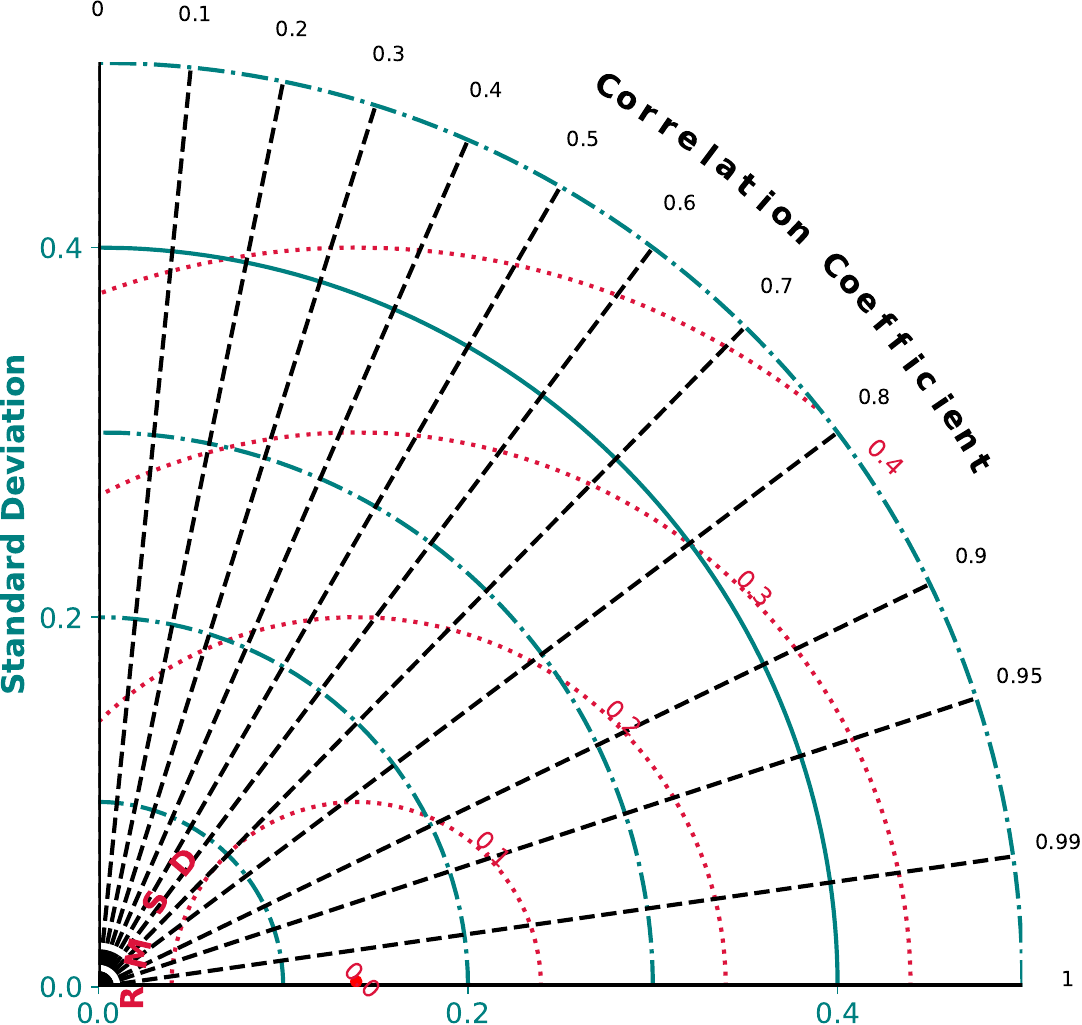}}
	\hfill
	\subfloat[Cp]{\includegraphics[width=0.3\textwidth]{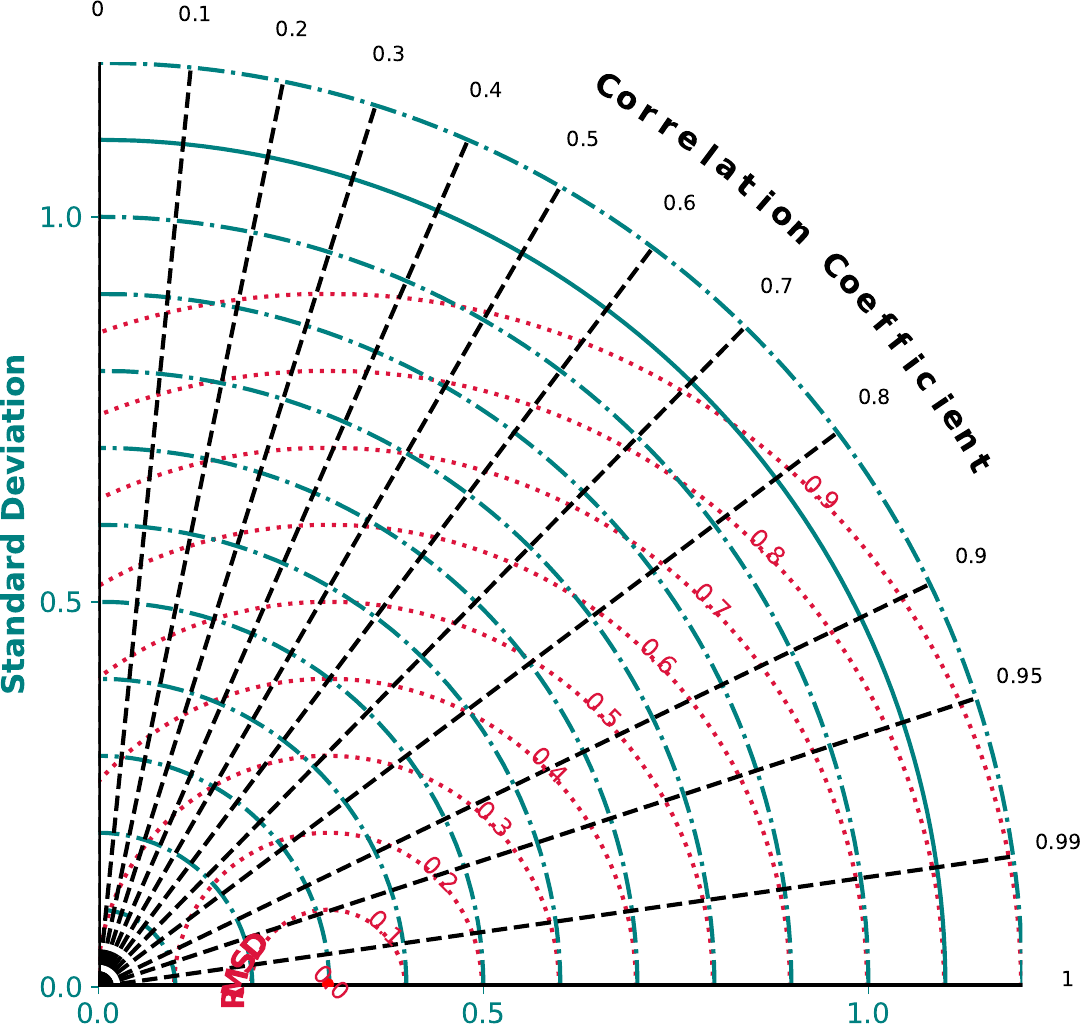}}
	\caption{Taylor diagram between PHengLEI and FU-CBAM-Net for NACA4424 Re=2000 AOA=5\degree.}
	\label{Taylor diagram naca4424 2000 5}
\end{figure*}

Figure \ref{Taylor diagram naca4424 2000 5} further utilizes a Taylor diagram to showcase the correlation coefficient, CRMSE, and standard deviation between the predicted values of the neural network model FU-CBAM-Net and the ground-truth values.
The test results from Fig. \ref{Taylor diagram naca4424 2000 5} reveal that for velocity components $u$, $v$, and pressure coefficient Cp, the correlation coefficients are all close to 1, and the CRMSE values are close to 0. 
This indicates a strong correlation and minimal error between the predicted results and the ground-truth values.

\begin{figure*}[!h]
	\begin{center}
		\includegraphics[width=1 \linewidth]{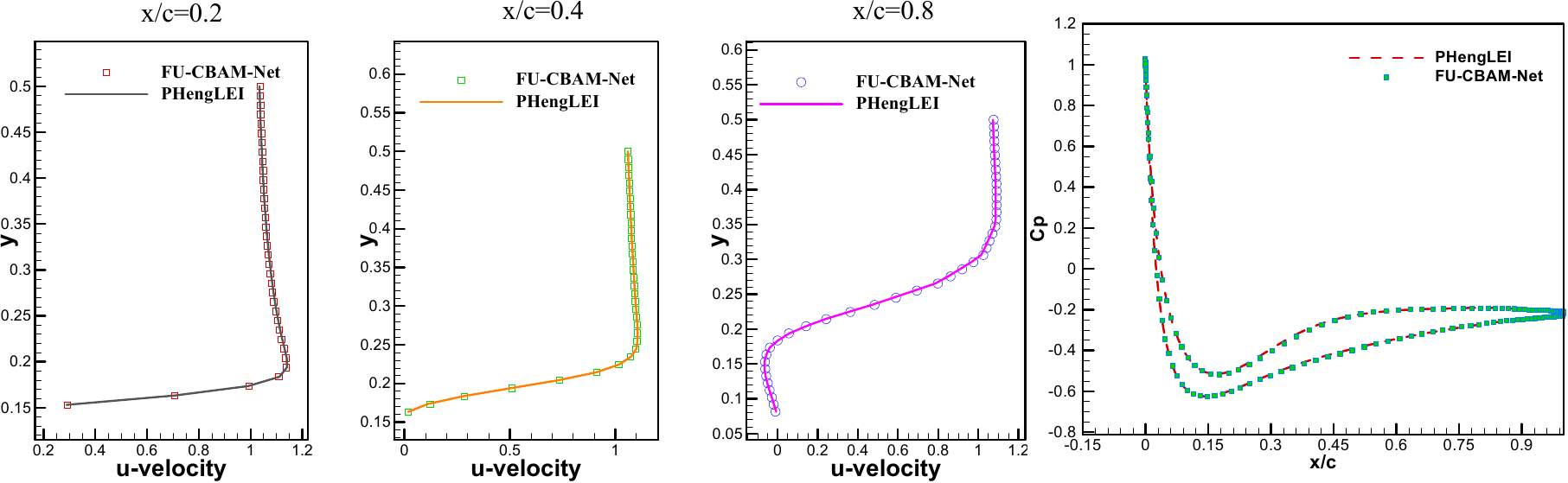}
	\end{center}  \vspace{-2mm}  
	\caption{{{Comparison between various velocity profiles and pressure coefficient for the reconstructed flow field and the ground-truth data. (a) velocity profiles. (b) Cp.}
	}} \label{naca4424 2000 5 velocity profiles} 
\end{figure*}

Figure \ref{naca4424 2000 5 velocity profiles} provides fitting curves of velocity profiles at various stations on the airfoil surface and fitting curves of pressure coefficient Cp between predicted values and ground-truth values.
From the test results, both the predictive curves of FU-CBAM-Net and the computed curves of PHengLEI show a strong fit, further indicating the high predictive accuracy achieved by the FU-CBAM-Net neural network model.
Here, only the test results for the NACA2415 Re=1600 AOA=3\degree and NACA4424 Re=2000 AOA=5\degree cases are provided. 
For additional test results, please refer to \ref{fu-cbame-net test results}.

Table \ref{FU-CBAM computational acceleration} further compares the computational times of PHengLEI and the FU-CBAM-Net neural network model for various test cases.
From the test results, it can be concluded that utilizing a well-trained neural network model achieves a three-order-of-magnitude acceleration compared to traditional CFD computational methods.
Additionally, compared to traditional CFD computation methods, the trained deep learning model has a smaller memory footprint, allowing for cross-platform development, easy portability, and rapid deployment.

\begin{table}[tb]
	\caption{FU-CBAM-Net and PHengLEI Computational Efficiency Comparison} \label{FU-CBAM computational acceleration}
	\begin{center}
		\begin{tabular}{clcc}
			\hline \hline
			\multicolumn{2}{l}{}                                                                                   & PHengLEI               & FU-CBAM-Net             \\ \hline
			\multicolumn{2}{c}{\multirow{2}{*}{\begin{tabular}[c]{@{}c@{}}NACA2415 \\ Re=1600 AOA=3 \degree\end{tabular}}} & \multirow{2}{*}{2244s} & \multirow{2}{*}{0.565s} \\
			\multicolumn{2}{c}{}                                                                                   &                        &                         \\ \hline
			\multicolumn{2}{c}{\begin{tabular}[c]{@{}c@{}}NACA4424\\ Re=2000 AOA=5\degree\end{tabular}}                   & 2293s                  & 0.582s                  \\ \hline
			\multicolumn{2}{c}{\begin{tabular}[c]{@{}c@{}}NACA2424\\ Re=1200 AOA=1\degree\end{tabular}}                   & 2237s                  & 0.578s                  \\ \hline
			\multicolumn{2}{c}{\begin{tabular}[c]{@{}c@{}}NACA0018\\ Re=1800 AOA=3\degree\end{tabular}}                   & 2281s                  & 0.573s                  \\ \hline \hline
		\end{tabular}
	\end{center} 
	\vspace{-1.5em}
\end{table} 

\subsection{Discussion of PHengLEI calculation results}

\begin{figure*}[!h] 
	\begin{center}
		\includegraphics[width=0.7 \linewidth]{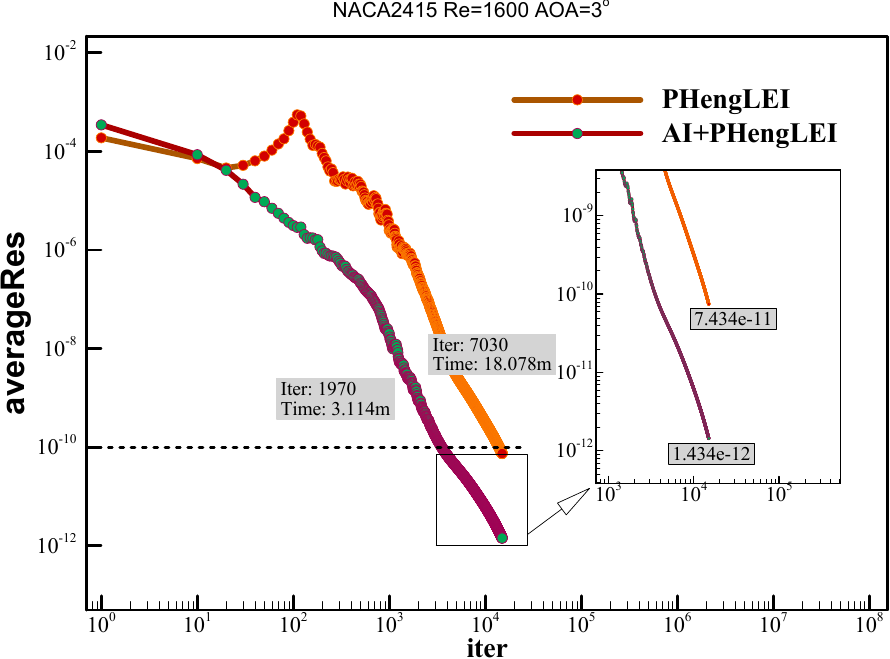}
	\end{center}  \vspace{-2mm}  
	\caption{{{Average residual convergence curves between the PHengLEI solver and the solver embedded with the FU-CBAM-Net neural network model for the NACA2415 Re=1600 AOA=3\degree.}
	}} \label{PHengLEI_res} 
\end{figure*}

\begin{figure*}[!h]
	\begin{center}
		\includegraphics[width=0.9 \linewidth]{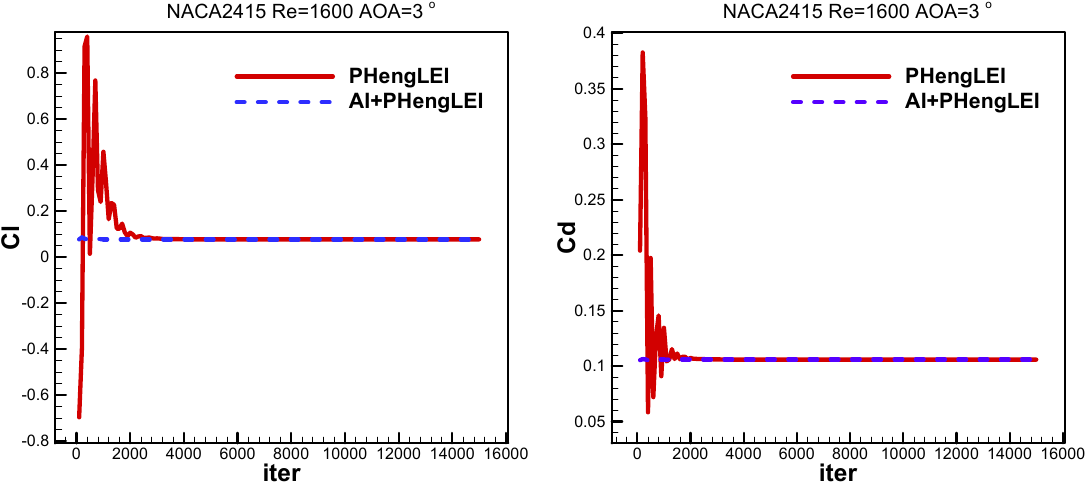}
	\end{center}  \vspace{-2mm}  
	\caption{{{Convergence curves for lift and drag during the iterative solving process for PHengLEI and AI+PHengLEI methods.}
	}} \label{naca2415_1600_3_cl_cd} 
\end{figure*}

Although deep learning-based methods for predicting flow fields have made some progress in recent years, researchers still consider them as opaque black-box models. 
The fundamental reason is the lack of reliable posterior credibility assessment strategies for the predicted results of flow fields.
For this purpose, we propose a novel credibility assessment method for the predicted results of neural network models. 
By embedding the flow field prediction results of the deep learning model into the PHengLEI solver, we aim to enhance the credibility of the flow field predictions. 
Additionally, incorporating converged solutions of the flow field into the solver can further accelerate the computational speed of the PHengLEI solver.
Here, we investigate the flow field posterior methods using the PHengLEI solver and the iterative solution acceleration efficiency using FU-CBAM-Net neural network, focusing on the NACA2415 Re=1600 AOA=3\degree and the NACA4424 Re=2000 AOA=5\degree cases as studied in Section \ref{flow field prediction results}.
Figure \ref{PHengLEI_res} presents the average residual convergence curves between the PHengLEI and the AI+PHengLEI (utilizing the predictions of FU-CBAM-Net as initial values for iterative solving of the NS equations with PHengLEI) for the NACA2415 Re=1600 AOA=3 \degree.
Here, considering e-10 as the residual convergence reference denoted as $F$, when PHengLEI converges to $F$, it takes 7030 iterations and 18.078 minutes. 
However, for AI+PHengLEI, convergence to $F$ takes 3.114 minutes and 1970 iterations.
Compared to PHengLEI, the AI+PHengLEI method reduces the number of iterations by 3.57 times and shortens the iteration time by 5.81 times.
Additionally, from the comparison curves of lift and drag in Fig. \ref{naca2415_1600_3_cl_cd}, it can be observed that traditional CFD methods exhibit varying degrees of oscillations in both lift and drag values during the initial stages of the iterative solving process.
However, in contrast, the AI+PHengLEI method shows rapid convergence in both lift and drag values, almost forming a straight line even during the initial stages of computation. 
These test results further indicate that the flow field solution obtained through FU-CBAM-Net prediction satisfies the NS equations.

\begin{figure*}[!h]
	\begin{center}
		\includegraphics[width=0.7 \linewidth]{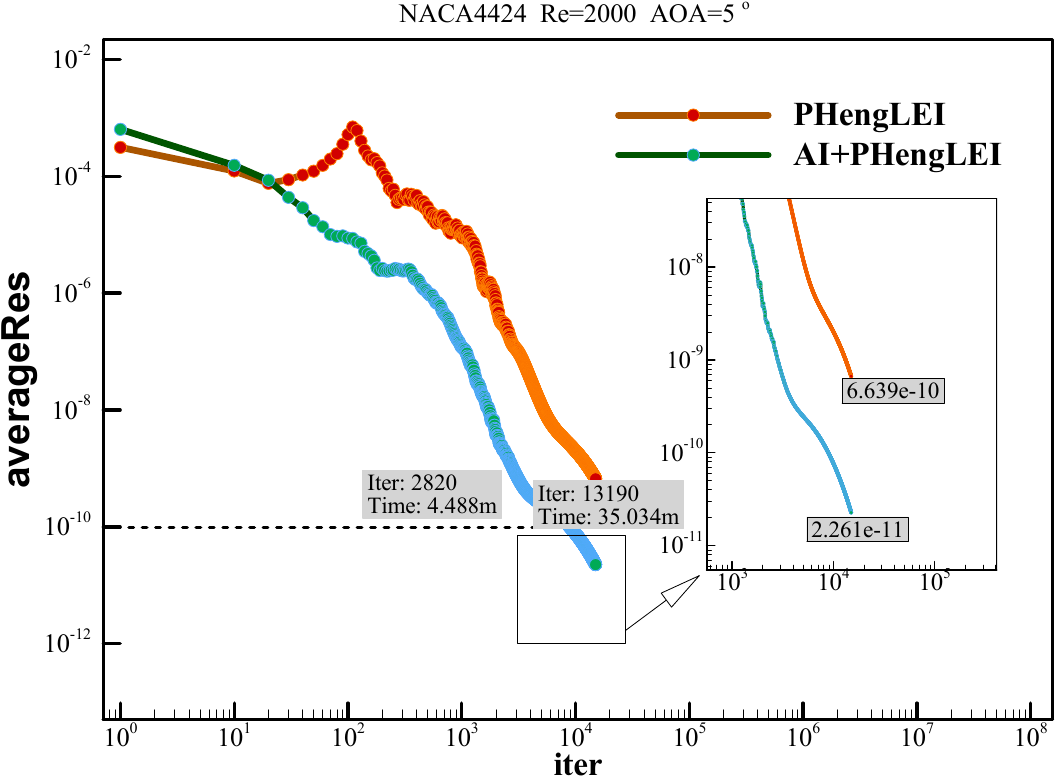}
	\end{center}  \vspace{-2mm}  
	\caption{{{Average residual convergence comparison curves between the PHengLEI solver and the solver embedded with the FU-CBAM-Net neural network model for the NACA4424 Re=2000 AOA=5\degree.}
	}} \label{naca4424_2000_5_averageRes} 
\end{figure*}

\begin{figure*}[!h]
	\begin{center}
		\includegraphics[width=0.9 \linewidth]{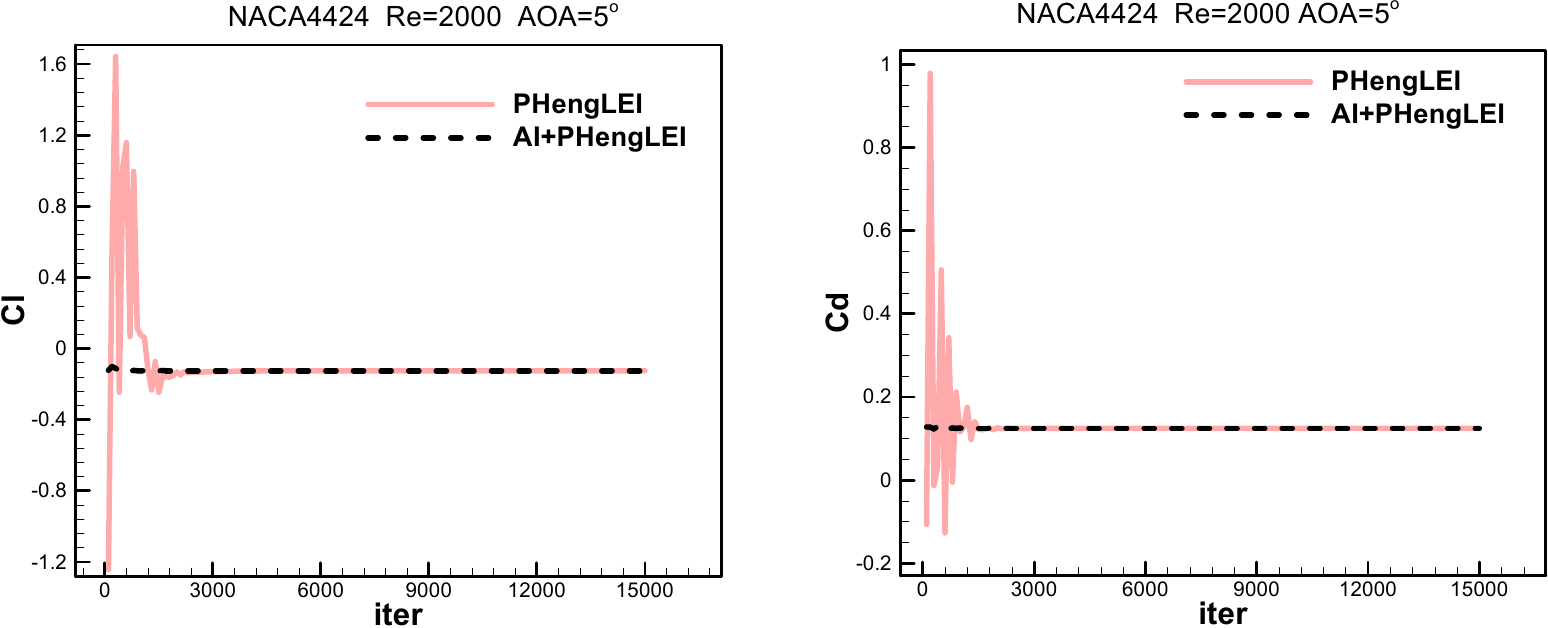}
	\end{center}  \vspace{-2mm}  
	\caption{{{Convergence curves for lift and drag during the iterative solving process for PHengLEI and AI+PHengLEI methods.}
	}} \label{naca4424_2000_5_cl_cd} 
\end{figure*}

To assess the generalization of the FU-CBAM-Net neural network model, Fig. \ref{naca4424_2000_5_averageRes} and Fig. \ref{naca4424_2000_5_cl_cd} presents the comparative results of residual convergence curves and lift/drag curves between PHengLEI and AI+PHengLEI methods during the flow field solution process for the NACA4424 Re=2000 AOA=5\degree.
As shown in Fig. \ref{naca4424_2000_5_averageRes}, similar to the test case for NACA2415 Re=1600 AOA=3 \degree. 
Here, $F$ is still used as the benchmark for residual convergence.
For the traditional PHengLEI solver, when converging to the benchmark $F$, it took 35.034 minutes and a total of 13190 iterations. 
However, for the accelerated PHengLEI solver with FU-CBAM-Net, it took 4.488 minutes with only 2820 iterations. 
Compared to the PHengLEI, the AI+PHengLEI method improved solving speed by 7.806 times and reduced the number of iterations by 4.677 times.
Moreover, under the same number of iterations, the residual of AI+PHengLEI is approximately one order of magnitude smaller than that of PHengLEI.
From the lift and drag convergence curves in Fig. \ref{naca4424_2000_5_cl_cd}, it's evident that the PHengLEI solver exhibits significant oscillations in lift and drag values during the initial iterative stages.
However, by embedding an artificial intelligence model within the traditional CFD solver, even during the initial iterations of the solver, convergence of values is achieved rapidly.

\section{Conclusions}

In this work, we proposed an enhanced UNet neural network model (FU-CBAM-Net) for rapid flow field prediction. 
Creatively, we integrated deep learning-based flow field prediction results into the PHengLEI software, establishing credibility validation for the predicted flow fields. 
FU-CBAM-Net's predictive accuracy and generalization were assessed across 24 NACA series airfoils under diverse operating conditions.
The primary conclusions of this study are as follows:

\begin{enumerate}
	\item Incorporating channel attention and spatial attention modules into the downsampling process of the conventional UNet neural network model significantly reduces model training loss and enhances the precision of flow field predictions.
	\item For individual airfoil flow fields, the prediction efficiency of the FU-CBAM-Net neural network model is three orders of magnitude faster than traditional CFD computational methods like PHengLEI.
	\item 
	Under given convergence criteria, embedding the prediction values of the FU-CBAM-Net deep learning model into the CFD computational process achieves an acceleration of at least five times and reduces iteration counts by over threefold. 
	Conversely, the aforementioned test results confirm the credibility of the flow field solution predicted by FU-CBAM-Net, adhering to the NS physical equations.
\end{enumerate}

Compared to most existing studies \cite{tang2021deep, sun2020surrogate}, this paper not only validates the potential of deep learning methods for accelerating airfoil physical field solutions but also demonstrates that deep learning models, through a more granular feature fusion strategy, can simulate complex separated flows.
This introduces an innovative method for traditional CFD technologies characterized by being time-consuming and memory-consuming.
In future work, we aim to extend our research into accelerating CFD solutions for three-dimensional complex flows using deep learning techniques.

\section *{CRediT authorship contribution statement}

$\mathbf{Kuijun \quad Zuo:}$ Data curation, Formal analysis, Investigation, Methodology, Software, Validation, Visualization, Writing-original draft, Writing-review \& editing.

$\mathbf{Zhengyin \quad Ye:}$ Funding acquisition, Supervision, Resources, Project administration. 

$\mathbf{Shuhui \quad Bu:}$ Methodology, Software, Writing-review \& editing.

$\mathbf{Xianxu \quad Yuan:}$ Funding acquisition, Resources, Supervision.

$\mathbf{Weiwei \quad Zhang:}$ Conceptualization, Methodology, Project administration, Supervision, Writing-review \& editing.

\section *{Acknowledgments}
This work was supported by the National Natural Science Foundation of China (Grant No. 12202470).

\section *{Declaration of competing interest}
The authors declare that they are have no known competing financial interests or personal relationships that could have appeared to influence the work reported in this paper.

\section *{} 

\bibliography{FU-CBAM-Net}

\newpage
\setcounter{figure}{0} 
\appendix 

\section{Taylor diagram} \label{Taylor diagram}

\begin{figure*}[!h]
	\begin{center}
		\includegraphics[width=0.4 \linewidth]{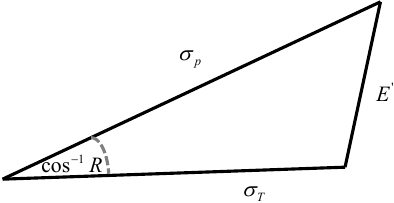}
	\end{center}  \vspace{-2mm}  
	\caption{{{Geometric relationship between the correlation coefficient, CRMSE, and standard deviation.}
	}} \label{Taylor} 
\end{figure*}

The Taylor diagram represents evaluation indicators such as the correlation coefficient, centered root-mean-square error (CRMSE), and standard deviation of a tested model on the same polar coordinate diagram. 
As shown in Fig. \ref{Taylor}, the three evaluation indicators satisfy the following cosine relationship:

\begin{equation} \label{eq:A1}
	E^{'2} = \sigma^2_p + \sigma^2_t - 2 \sigma^2_p \sigma^2_t R
\end{equation}

In Eq. \eqref{eq:A1}, $\sigma_p$ and $\sigma_t$ are the standard deviations of the predicted values and the ground-truth values, respectively.
The calculation formula is as follows:
\begin{equation}
	 \sigma_{\kappa} = \left[\frac{1}{N} \sum_{i=1}^N\left(\kappa_i-\bar{\kappa}\right)^2\right]^{\frac{1}{2}}, \kappa = p, t.
\end{equation}

In the above formula, $\bar{\kappa}$ represents the mean values of either the ground-truth or predicted values.
Moreover, in Eq. \eqref{eq:A1}, $R$ and $E$ represent the correlation coefficient and CRMSE, respectively.
The calculation formula for the correlation coefficient $R$ is given as follows:

\begin{equation}
	R=\frac{\frac{1}{N} \sum_{i=1}^N\left(p_i-\bar{p}\right)\left(t_i-\bar{t}\right)}{\sigma_p \sigma_t}
\end{equation}
where $\bar{p}$ and $\bar{t}$ are the mean values of predicted values $p$ and ground-truth data $t$.

Additionally, the root-mean-square error (RMSE) is defined as:

\begin{equation}
	E = \left[ \frac{1}{N}\sum_{i=1}^{N}(p_i - t_i)^2 \right]^{\frac{1}{2}}
\end{equation}

In literature \cite{taylor2001summarizing}, the RMSE is decomposed into two parts, overall bias $\bar{E}$ and CRMSE $E^{'}$.
These two parameters are defined as:

\begin{equation} \label{eq:A3}
\left\{
\begin{aligned}
	 \bar{E} &= \bar{p} - \bar{t}, \\
	 E^{\prime}&=\left\{\frac{1}{N} \sum_{i=1}^N\left[\left(p_i-\bar{p}\right)-\left(t_i-\bar{t}\right)\right]^2\right\}^{1 / 2}
\end{aligned}
\right.	
\end{equation}

Additionally, the amalgamation of these two elements results in a quadratic accumulation, yielding the comprehensive mean square difference:

\begin{equation}
	E^2=\bar{E}^2+E^{\prime 2}
\end{equation}

\newpage
\setcounter{figure}{0} 

\section{FU-CBAM-Net flow field prediction results} \label{fu-cbame-net test results}

\begin{figure*}[!h]
	\begin{center}
		\includegraphics[width=0.8 \linewidth]{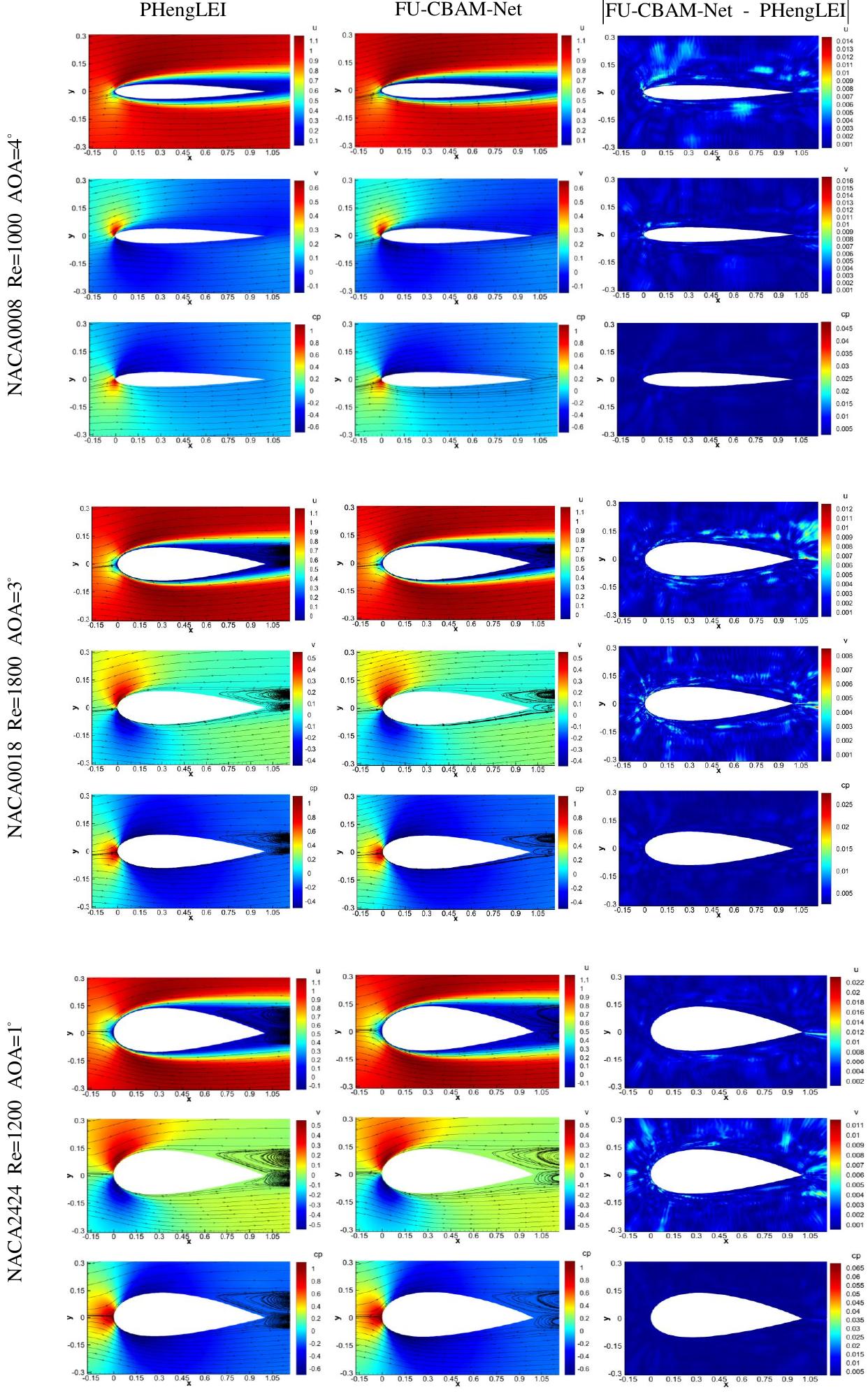}
	\end{center}  \vspace{-2mm}  
	\caption{{{Predicted results of different airfoil flow fields.}
	}} \label{appendix B} 
\end{figure*}

\end{document}